\documentclass[review,authoryear]{elsarticle}
\usepackage{amsmath}
\usepackage{marginnote}
\usepackage{epstopdf}
\usepackage{parskip}
\usepackage{natbib}
\usepackage{mathtools}
\usepackage{graphicx}
\usepackage{epstopdf}
\usepackage{caption}
\usepackage{amssymb}
\usepackage{listings}
\usepackage{float}
\usepackage[position=top]{subcaption}
\usepackage{enumerate}	
\usepackage{esint} 
\usepackage{paralist} 
\usepackage{placeins} 
\usepackage{setspace} 
\usepackage{wasysym} 
\usepackage{xfrac} 
\usepackage{color} 
\usepackage{textcomp}
\usepackage{nomencl}
\usepackage{siunitx}
\usepackage{booktabs}

\newcommand{\Reyn}{\operatorname{\mathit{R\kern-.04em e}}}

\usepackage{tikz}
\usetikzlibrary{patterns}

\newcommand{\reddashed}{\raisebox{2pt}{\tikz{\draw[red,dashed,line width=0.9pt](0,0) -- (5mm,0);}}}
\newcommand{\redsolid}{\raisebox{2pt}{\tikz{\draw[red,solid,line width=0.9pt](0,0) -- (5mm,0);}}}
\newcommand{\bluedashed}{\raisebox{2pt}{\tikz{\draw[blue,dashed,line width=0.9pt](0,0) -- (5mm,0);}}}
\newcommand{\bluesolid}{\raisebox{2pt}{\tikz{\draw[blue,solid,line width=0.9pt](0,0) -- (5mm,0);}}}
\newcommand{\orangesolid}{\raisebox{2pt}{\tikz{\draw[orange,solid,line width=0.9pt](0,0) -- (5mm,0);}}}
\newcommand{\greensolid}{\raisebox{2pt}{\tikz{\draw[green,solid,line width=0.9pt](0,0) -- (5mm,0);}}}
\newcommand{\cyansolid}{\raisebox{2pt}{\tikz{\draw[cyan,solid,line width=0.9pt](0,0) -- (5mm,0);}}}
\newcommand{\Exp}{\raisebox{2pt}{\tikz{\draw (0,0) circle (2pt);}}}

\journal{Renewable Energy Journal}
\begin{document}
\begin{frontmatter}
\title{Turbulence--resolving simulations of wind turbine wakes}
\author[1]{Georgios Deskos}
\ead{g.deskos14@imperial.ac.uk}
\author[2]{Sylvain Laizet}
\author[1]{Matthew D. Piggott}
\address[1]{Department of Earth Science and Engineering, Imperial College London, London, SW7 2AZ, UK}
\address[2]{Department of Aeronautics, Imperial College London, London, SW7 2AZ, UK}
\date{\today}
\begin{abstract}
Turbulence-resolving simulations of wind turbine wakes are presented using a high--order flow solver combined with both a standard and a novel dynamic implicit spectral 
vanishing viscosity (iSVV and dynamic iSVV) model to account for subgrid-scale (SGS) stresses. The numerical solutions are compared against wind 
tunnel measurements, which include mean velocity and turbulent intensity profiles, as well as integral rotor quantities such as power and 
thrust coefficients. For the standard (also termed static) case the magnitude of the spectral vanishing viscosity is selected via a heuristic analysis of the wake 
statistics, while in the case of the dynamic model the magnitude is adjusted both in space and time at each time step. The study focuses on 
examining the ability of the two approaches, standard (static) and dynamic, to accurately capture the wake features, both qualitatively and 
quantitatively. The results suggest that the static method can become over-dissipative when the magnitude of the spectral viscosity 
is increased, while the dynamic approach which adjusts the magnitude of dissipation locally is shown to be more appropriate for 
a non--homogeneous flow such that of a wind turbine wake.
\end{abstract}
\end{frontmatter}
\section{Introduction}
Understanding and modelling wake--turbine and wake--wake interactions within wind farms have been recognised as one of the long--term challenges in wind energy research 
\citep{vanKuikEtAl2016}. Existing wind farm wake models vary from low-fidelity empirical and semi-empirical approaches \citep{Jensen1983,Ainslie1988} to more 
sophisticated high--fidelity large-eddy simulations (LES) where the turbines are parametrized using either an actuator disc \citep{JimenezEtAl2007} or an 
actuator line approach \citep{TroldborgEtAl2010,TroldborgEtAl2011}. The latter, despite being the most 
computationally expensive method, has recently attracted the interest of the research community and has been applied to the 
study of many utility-scale wind farms \citep{Ivanell2009,ChurchfieldEtAl2012,NilssonEtAl2015}. 
This is primarily due to the increasing computational capacity of modern high--performance computer (HPC) platforms,
but also as a result of the ability of LES to resolve turbine wakes to significantly finer spatial and 
temporal scales and thus provide deep insights into the complexities of wake dynamics. It is evident that as we move towards longer and more realistic wake 
simulations we need to optimise the use of existing and future HPC resources in order to obtain as much information as possible from a particular problem size (e.g. 
expressed in terms of computational degree of freedom count). In the context of LES, we argue that this entails two factors: first the need for 
subgrid--scale (SGS) models that do not act in an over--dissipative manner so that larger coherent structures of turbulence can be retained, and second 
the use of high--order schemes with spectral or ``spectral-like'' accuracy which can capture more flow field details with the same degree of freedom count.

Starting with the former, the development and validation of SGS models has long been an active area of research, with numerous models having been suggested and 
applied to turbine wakes. A general review of the various SGS models can be found in \cite{MeneveauKatz2000} and \cite{Sagaut2006}, whereas a more wind turbine 
wakes specific review (including their interactions with the atmospheric boundary layer) was recently presented by \cite{MehtaEtAl2014}. \cite{SarlakEtAl2015} 
also studied the role of SGS modelling in predicting wake statistics in wind tunnel test configurations by examining a number of explicit SGS models, including the standard 
Smagorinsky (SS) model \citep{Smagorinsky1963}, the dynamic Smagorinsky (DS) model \citep{GermanoEtAl1991}, as well as the dynamic mixed model of \cite{ZangEtAl1993}. 
The authors concluded that the selection of the particular SGS modelling approach is not a primary factor in determining the overall accuracy of the obtained results. Instead 
they argued that a good turbine parametrisation (in their particular case the use of sufficient actuator line resolution) is more important than the SGS modelling 
choice, and that implicit models which employ numerically dissipative upwind schemes can provide equally good results. \cite{MehtaEtAl2014} also concluded that if an adequate spatial resolution is used, the impact of the SGS model choice is nullified.

Employing high--order methods with spectral or ``spectral--like'' accuracy has also been a common practice in LES of wind turbine wakes and wind farms. 
For instance, \cite{CalafEtAl2010} and \cite{LuPorte-Agel2011} used formulations which employ spectral schemes in the lateral directions, together with periodic 
boundary conditions, and energy conserving finite--difference schemes in the vertical direction. The use of periodic boundaries for spectral methods is imperative 
and limits the method to simple flow configurations. More recently, first \cite{PeetEtAl2013} and latterly \cite{KleusbergEtAl2017} developed turbine 
parametrisation models within the hp/spectral element solver Nek5000 \citep{nek5000-web-page}. These 
formulations exhibit a number of advantages compared to the spectral schemes just mentioned. Not only do they provide 
higher accuracy in comparison to more conventional second-- or fourth--order based models, they also allow 
complex geometries to be used in the simulations of wind farms 
(e.g. uneven terrain). Nevertheless, when spectral or ``spectral--like'' higher--order methods 
are used, a certain amount of numerical dissipation is still required to stabilise the numerical solution.
In the case of spectral methods, a standard procedure is to truncate the smaller scales by 
using the so-called \num{3/2} rule \citep{Orszag1970}. On the other hand, when ``spectral--like'' accurate 
methods are used, this stabilisation is achieved either by employing a spectral--vanishing 
viscosity (SVV) operator \citep{Tadmor1989,KaramanosKarniadakis2000}, to selectively add 
dissipation to the near grid cut--off scales, or by applying a filter-based stabilisation to 
remove energy from the highest modes \citep{FischerMullen2001}. Higher--order ``upwind-biased'' numerical schemes 
can also be used \citep{RaiMoin1991,MengaldoEtAl2017} to stabilise the numerical solution. All these methods can deliver 
enhanced eddy viscosities and therefore can arguably be used as SGS models for LES. The former approach that uses a SVV operator for LES of turbulent flows has been demonstrated by 
\cite{KaramanosKarniadakis2000,KirbyKarniadakis2002,Pasquetti2005,Pasquetti2006,PasquettiEtAl2008,SeveracSerre2007,KoalEtAl2012} and \cite{LombardEtAl2016}. It should be noted that the action of the SVV operator is strictly 
dissipative, thus the SVV--LES formulation may also be regarded and referred to as an implicit LES (iLES) technique 
\citep{Sagaut2006}, although its formulation and action is fundamentally different from that of the MILES 
(Monotonically Integrated LES) approach \citep{GrinsteinFureby2004}, for example. More recently an implicit SVV 
model based on the manipulation of the second derivative of high--order compact finite--difference schemes was 
introduced by \cite{LamballaisEtAl2011} and applied in different flow configurations by \cite{DairayEtAl2014,DairayEtAl2017} and \cite{IoannouLaizet2018}. \cite{DairayEtAl2017} also determined the magnitude of the required SVV in the context of isotropic turbulence using Pao's equilibrium energy spectrum. 
Unfortunately, the proposed methodology cannot easily be extended to non-homogeneous turbulent flows and 
one can only select a value based on a trial-and-error basis. However, even an optimum selected value (on the 
basis of better agreement with reference data) will introduce the same amount of dissipation everywhere in the computational domain without taking into account local 
flow features (e.g. local turbulent kinetic energy or the filtered strain-rate tensor) or non-homogeneous and transitional regions. To overcome this problem, a dynamic version of 
the SVV was proposed by \cite{KirbyKarniadakis2002} and successfully applied to turbulent channel flow.

Applying the SVV approach to wind turbine wake simulations will inevitably require a careful consideration of the particular flow structures. In particular, a 

turbine wake exhibits three distinct regions \citep{VermeerEtAl2003,SanderseEtAl2011}: the near-wake field which extends up to approximately a diameter 
downstream the rotor and is dominated by the dynamics of the blade tip vortices (laminar region), the transitional (or merging zone) in which the large 
well-structured vortices begin to de-stabilise and eventually break up into smaller eddies (transitional region), and finally the far-wake field in which the 
wake flow tends -- in the absence of an external forcing (e.g. an atmospheric boundary layer) -- towards a self-similar wake (turbulent region). Given this
inherent non-homogeneity of the flow field, it is expected that different levels of dissipation will need 
to be utilised. Considering all these, we pose the following questions regarding the two SVV (static and dynamic) approaches:
\begin{itemize}
\item Is the static SVV--LES appropriate for LES of high-Reynolds number wind turbine wakes?
\item How sensitive are results to the selection of the static SVV magnitude?
\item Can the dynamic SVV approach yield better results compared to the static approach, on the basis that dissipation will be scaled locally? 
\end{itemize}
To answer these questions we make use of sixth-order compact finite--difference schemes, the implicit SVV approach of 
\cite{DairayEtAl2017} and a newly developed implicit dynamic SVV model which scales the amount of 
local dissipation with the magnitude of the strain-rate tensor, by invoking the linearity of the dissipation operator developed by \cite{LamballaisEtAl2011}. 
As reference data against which we compare our numerical simulations we select the ``blind tests" database of 
\cite{KrogstadEriksen2013} and \cite{PierellaEtAl2014}. These wind tunnel measurements have been 
used by a number of researchers \citep{Sarlak2014,SorensenEtAl2015,KleusbergEtAl2017}
and therefore represent an excellent benchmark case to compare against previous as well as future studies. To parametrise the wind turbines we make use of the 
actuator line technique \citep{SorensenShen2002} based on a turbine parametrisation which has been previously shown to better capture the key features of the 
near wake field (e.g. tip vortices).

This remainder of this paper begins with a presentation of the numerical discretisation techniques employed, including details on the static and dynamic SVV methods and a short description of the actuator line turbine parametrisation used in this work. 
Subsequently, the wind tunnel tests are briefly described in section \ref{sec:SimSetup} and compared with the numerical model solutions in section 
\ref{sec:Comparison}. Additional qualitative and quantitative comparisons and observations for the obtained wake solutions and their sensitivity to input 
parameters are provided in section \ref{sec:AdditionalComparison} and further discussed in section \ref{sec:Discussion}.

\section{Numerical solver}\label{sec:Theory}
To resolve turbine wakes we consider a modified version of the incompressible Navier-Stokes equations
\begin{subequations}
\begin{align}
\frac{\partial u_i}{\partial t}+ \frac{1}{2}\bigg(u_j\frac{\partial u_i}{\partial x_j}+
\frac{\partial u_i u_j}{\partial x_j}\bigg)&=-\frac{1}{\rho}\frac{\partial p}{\partial x_i}+
\nu\mathcal{D}(u_i)+\frac{F_i^T}{\rho},\label{eq:NS_mom}\\
\frac{\partial u_i}{\partial x_i}&=0,\label{eq:NS_cont}
\end{align}
\label{eq:NS}
\end{subequations}
where $u_i$ for $i\in\{x,y,z\}$ represents the velocity component in the $(x,y,z)$ directions respectively, 
$p$ is the pressure field, $\nu\mathcal{D}(u_i)$ is the dissipation term, $F^T$ is the turbine actuator 
forcing and $\nu$, $\rho$ are the kinematic viscosity and the fluid density respectively. 

The dissipation operator $\mathcal{D}$ is evaluated by taking advantage of the discretisation error of the second derivatives when high--order compact finite--differences schemes are used, as will be described below. It is designed to mimic the combined effects of conventional viscous and SVV operators and can be expressed as 
\begin{equation}
\nu \mathcal{D}(u_i) \approx \nu\frac{\partial^2 u_i}{\partial x_j\partial x_j}+\frac{\nu_0}{\nu}\frac{\partial}{\partial x_j}\bigg[ \mathcal{Q}_k * \frac{\partial u_i}{\partial x_j}\bigg],
\end{equation}
where the star $(*)$ denotes the convolution process that would need to be used in an explicit SVV formulation. $\nu_0$ is a user-defined parameter which determines the magnitude of the SVV operator, and $\mathcal{Q}_k$ is the wavenumber-dependent SVV kernel defined by \cite{KaramanosKarniadakis2000}: 
\begin{equation}
\mathcal{Q}_k= \begin{cases} 
    0 & \text{if  } k < 0.3k_c \\
   \exp\bigg[-\bigg(\frac{k_c-k}{0.3k_c-k}\bigg)^2\bigg] & \text{if  } 0.3k_c \leq k \leq k_c,
 \end{cases}
 \label{eq:ExplicitSVV}
\end{equation}
where $k$ is the velocity-based wavenumber and $k_c=\pi/\Delta x$ is the cut-off wavenumber associated with the mesh size $\Delta x$.

In this work our particular {\it implicit} SVV formulation does not require us to implement any of the explicit SVV processes (e.g. convolution) or even explicitly define
the SVV kernel in the model. Instead, an SVV--like kernel is obtained by controlling the numerical dissipation at the finer spatial scales using the discretisation error of the schemes. 
Further details on the implicit model are provided for both the standard and the dynamic cases in sections \ref{subsec:SVV} and \ref{subsec:DSVV} below. 

\begin{figure}[t]
\centering
\includegraphics[width=0.9\linewidth]{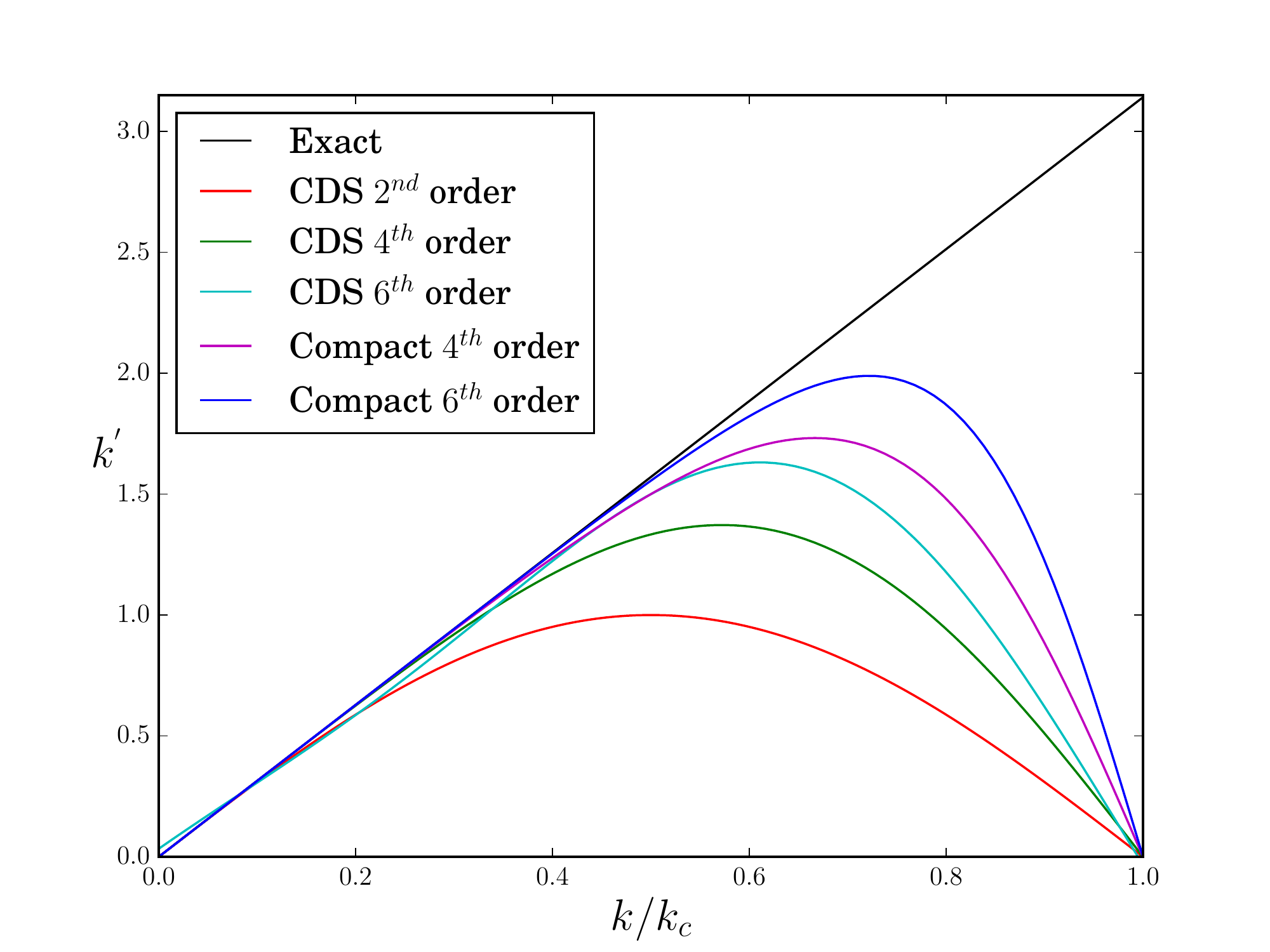}
\caption{Plot of the modified wavenumber $k'$ vs the theoretical wavenumber $k$.}
\label{fig:CDFirstDerError}
\end{figure}

To solve 
equations \eqref{eq:NS_mom} and \eqref{eq:NS_cont}, the high--order finite-difference flow solver \texttt{Incompact3d} \citep{LaizetLamballais2009,LaizetLi2011} is 
used. It is based on a Cartesian mesh in a half-staggered arrangement (the same mesh is used for the three velocity components $u_x,u_y,u_z$, with a different mesh used for pressure $p$), sixth--order compact finite--differences schemes \citep{Lele1992} for the spatial discretisation, an explicit third--order Runge--Kutta method for the time advancement, and a direct spectral solver for the Poisson equation. The staggered mesh arrangement is essential in order to avoid spurious pressure oscillations on the mesh, while the compact sixth-order finite-difference schemes are essential to obtain the required ``spectral--like'' behaviour. To appreciate the importance of compact sixth--order accurate schemes, we briefly present in figure \ref{fig:CDFirstDerError} the modified wave number of the first derivative and compare it to other standard central difference schemes (CDS) commonly used in energy conservative numerical solvers.  
The modified wave number $k'$ is obtained by conducting a Fourier analysis of the discrete first derivative of any velocity component $u_j$, $j \in \{x,y,z\}$,
to obtain $\widehat{du_j/dx}=i k' \hat{u}_j$, where $i=\sqrt{-1}$ is the imaginary unit and $\widehat{(\cdots)}$ denotes the Fourier transform of the function. From figure \ref{fig:CDFirstDerError} we may observe that from all the presented schemes the sixth--order compact finite--differences scheme better matches the exact solution ($k'=k$) as it remains accurate for a relative wavenumber $k/k_c$ of up to around $2/3$. We refer to this scheme as a ``spectral--like'' accurate  scheme due to its ability to capture finer spatial scales. We take care of the other discretisation errors such as the aliasing error appearing in the numerical solution  by using the skew-symmetric form of the convective term as suggested by \cite{KravchenkoMoin1997}. Finally, \texttt{Incompact3d} is parallelised with the aid of MPI and an efficient 2D pencil domain decomposition approach \citep{LaizetLi2011}. The turbine momentum source, $F^T$, which is computed at each time step, uses a standard actuator line implementation, a description of which is briefly presented in section \ref{subsec:ALM}.

\subsection{The ``standard'' (static) iSVV approach}\label{subsec:SVV}
\begin{figure}[t]
\centering
\includegraphics[width=0.9\linewidth]{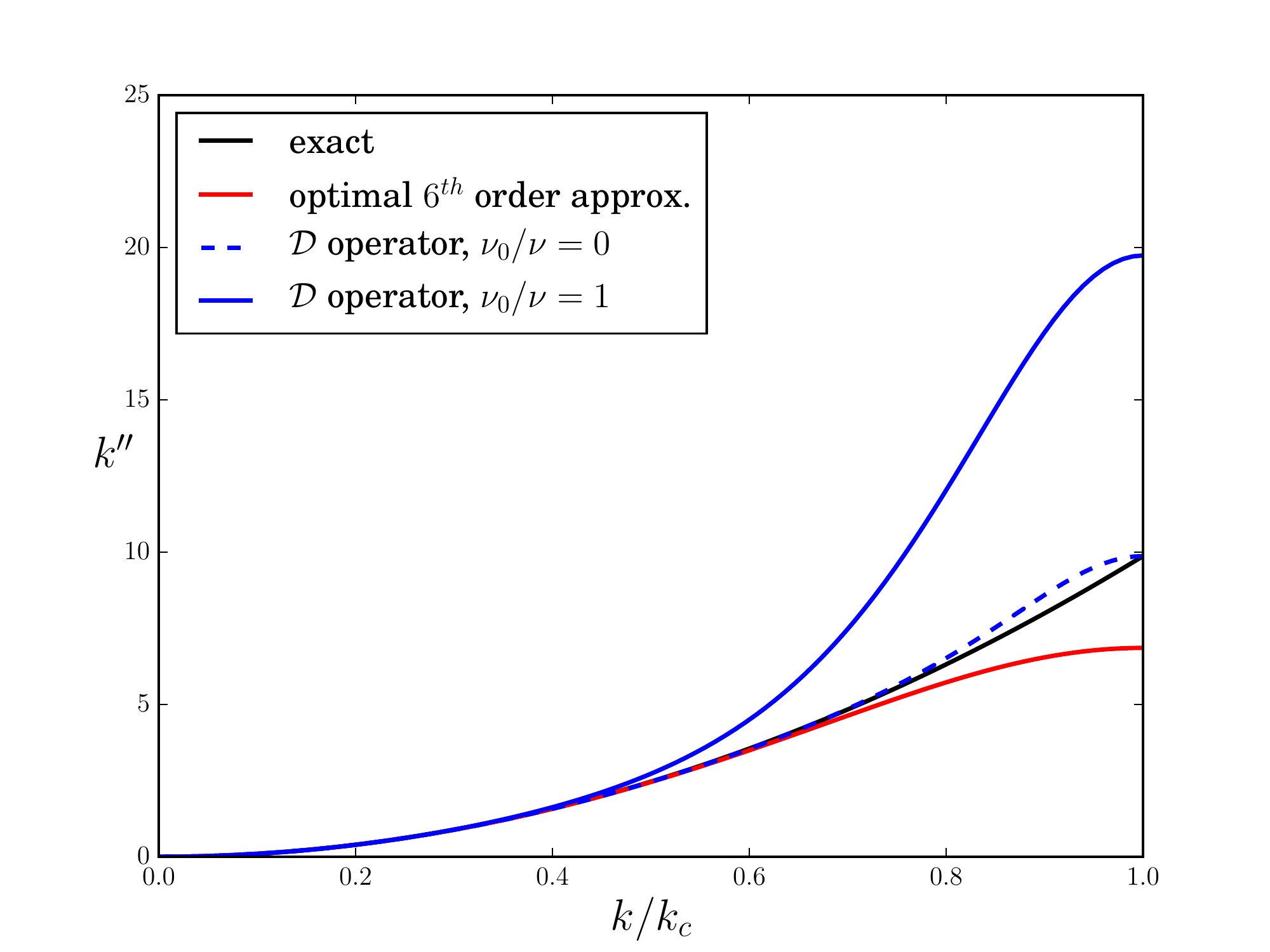}
\caption{Modified wavenumber of the second derivatives plotted against the normalised wavenumber $k/k_c$. The four plotted curves are the exact solution, the optimal \num{6}$^{th}$ order accurate scheme, and the dissipation operator $\mathcal{D}$ for the values $\nu_0/\nu=\,$\num{0} and $\nu_0/\nu=\,$\num{1}.}
\label{fig:CDSecDerError}
\end{figure}

Since the dissipation operator $\mathcal{D}$ is a non-standard one in the literature, it is instructive to discuss its implementation in greater 
detail. Starting with a general representation of compact finite-differences schemes, the second derivative can be described through a 3-9 stencil formulation:
\begin{equation}
    \begin{split}
	\alpha f''_{i-1}+f''_{i}+\alpha f''_{i+1}
     &=
     a\frac{f_{i+1}-2f_i+f_{i-1}}{\Delta x^2}
     + b\frac{f_{i+2}-2f_i+f_{i-2}}{4\Delta x^2} \\
     &+ c\frac{f_{i+3}-2f_i+f_{i-3}}{9\Delta x^2}
 	+ d\frac{f_{i+4}-2f_i+f_{i-4}}{16\Delta x^2},
    \end{split}
\label{eq:CompactSecondDerivative}
\end{equation}
where $f_i=f(x_i)$ and $f''_i=f''(x_i)$ are the values of the function $f(x)$ at the 
nodes $x_i=(i-1)\Delta x$ and where $\Delta x$ is a uniform mesh spacing. The five 
coefficients $(\alpha,a,b,c,d)$ can be chosen in such a way so as to ensure up to tenth-order 
accuracy, by satisfying exactly the following five relations
\begin{subequations} 
\begin{align} 
a + b + c  +d=1 + 2 \alpha  \ \ & \text{(second order)} \label{eq:CompFirst2Condition}\\ 
a + 2^2 b + 3^2 c +4^2 d= \frac{4!}{2!} \alpha  \ \ & \text{(fourth order)}
\label{eq:CompFirst4Condition}\\ 
a + 2^4 b + 3^4 c +4^4 d = \frac{6!}{4!} \alpha  \ \ & \text{(sixth order)}\label{eq:CompFirst6Condition}\\ 
a + 2^6 b + 3^6 c +4^6 d= \frac{8!}{6!} \alpha \ \ & \text{(eight order)}\label{eq:CompFirst8Condition}\\
a + 2^8 b + 3^8 c +4^8 d= \frac{10!}{8!} \alpha \ \ & \text{(tenth order).}\label{eq:CompFirst10Condition} 
\end{align} 
\end{subequations} 
Requiring only sixth--order spatial accuracy, the last two equations \eqref{eq:CompFirst8Condition} and \eqref{eq:CompFirst10Condition} do not need to be
satisfied and therefore two out of the five coefficients can be chosen freely. \cite{Lele1992} showed that by choosing ($\alpha, a, b, c, d\,$)=(\num{2/11}, 
\num{12/11}, \num{3/11}, \num{0}, \num{0}) an ``optimal'' sixth--order accurate scheme with spectral-like behaviour can be obtained. To better appreciate the 
spectral behaviour of this ``optimal'' sixth--order scheme we may again present the modified (effective) wavenumber of the discrete second derivative and compare the modified (effective) second derivative wavenumber $k''$ to the exact solution $k^2$ (see figure \ref{fig:CDSecDerError}). Here we also present an expression of the modified second derivative wavenumber in terms of the five coefficients $(\alpha,a,b,c,d)$ associated with the compact scheme:
\begin{equation}
k''(k) = 
\dfrac{\splitdfrac{ 2a[1-\cos(k \Delta x)]+\frac{b}{2}[1-\cos(2 k \Delta x)]+}
{\frac{2c}{9}[1-\cos(3 k \Delta x)]+\frac{d}{8}[1-\cos(4 k \Delta x)]}}{\Delta x^2[1+2\alpha\cos(k \Delta x)]}.
\label{eq:ModifiedWaveNumber}
\end{equation}
Again, an excellent agreement can be observed up to approximately $k/k_c=2/3$. The rightmost region, $k/k_c>2/3$ (region of small scales), is said to exhibit 
under-dissipative behaviour for the second derivative due to the inability of the ``optimal'' sixth--order scheme to effectively remove the near-grid-size scales when considered in the context 
of direct numerical simulations (DNS). 
On the other hand, the ability to freely choose two of the coefficients in the discretisation presents 
a unique opportunity to add two constraints (equations) in an effort to add extra numerical
dissipation and turn the discrete second derivative approximation into an SVV-like operator. 
The two additional constraints are chosen in such a way so that the modified wavenumber $k''$ at 
$k=k_m=2/3 k_c$ and $k=k_c$ should match the respective wavenumber of a combined viscous term/SVV operator kernel:
\begin{subequations}
\begin{align}
k''(k_c)&=\bigg(1+\frac{\nu_0}{\nu}\bigg)k_c^2, \label{eq:nu_at_kc}\\
k''(2 k_c/3)&=\bigg(1+0.437\frac{\nu_0}{\nu}\bigg)k_m^2\label{eq:nu_at_2/3kc}.
\end{align}
\end{subequations}
Inherently, by setting $\nu_0/\nu=\,$\num{0}, the additional numerical dissipation effects (the ones that mimic the SVV operator) will be nullified and therefore a nearly exact
solution may be obtained (see fig \ref{fig:CDSecDerError}). Nevertheless, equations \eqref{eq:nu_at_kc}--\eqref{eq:nu_at_2/3kc} together with the 
original equations for sixth--order accuracy \eqref{eq:CompFirst2Condition}--\eqref{eq:CompFirst6Condition} make the 
system complete and the five coefficients can now be expressed in terms of $k''_m=k''(2 k_c/3)$ and $k''_c=k''(k_c)$ as
\begin{subequations}
\begin{align}
\alpha&=\frac{1}{2}-\frac{320k''_m \Delta x^2-1296}{405k''_c\Delta x^2 -640 k''_m\Delta x^2+144} \\
a&=-\frac{4329k''_c\Delta x^2/8 -32 k''_m \Delta x^2-140k''_c\Delta x^2+286}{405k''_c\Delta x^2-640k''_m\Delta x^2 +144}\\
b&= \frac{2115k''_c\Delta x^2 -1792 k''_m \Delta x^2-280k''_c\Delta x^2+1328}{405k''_c\Delta x^2-640k''_m\Delta x^2 +144}\\
c&=-\frac{7695k''_c\Delta x^2/8 +288 k''_m \Delta x^2-180k''_c\Delta x^2-2574}{405k''_c\Delta x^2-640k''_m\Delta x^2 +144}\\
d&=\frac{198k''_c\Delta x^2 +128 k''_m \Delta x^2-40k''_c\Delta x^2-736}{405k''_c\Delta x^2-640k''_m\Delta x^2 +144}.
\end{align}
\end{subequations}
The newly obtained operator is expected to exhibit enhanced levels of numerical dissipation for the higher wavenumbers due to the imposed conditions (equations \eqref{eq:nu_at_kc}--\eqref{eq:nu_at_2/3kc}). 
However, to appreciate the effective spectral viscosity we may assume that the modified wave number will act as any other spectral eddy viscosity model \citep{Kraichnan1976,CholletLesieur1981} and therefore we may argue that in the spectral space
\begin{equation}
-[\nu + \nu_s(k,k_c)]k^2 \hat{u}_k = -\nu k'' \hat{u}_k.
\label{eq:spectral_viscosity_equivalence}
\end{equation}
By re-arranging equation \eqref{eq:spectral_viscosity_equivalence}, we compute the implicit (or associated) spectral viscosity as
\begin{equation}
\frac{\nu_s(k,k_c)}{\nu}=\frac{k''-k^2}{k^2}.
\label{eq:AssociatedSpectralViscosity}
\end{equation}
\begin{figure*}[!ht]
\centering
\begin{subfigure}[t]{0.45\textwidth}
\centering
\includegraphics[width=\linewidth]{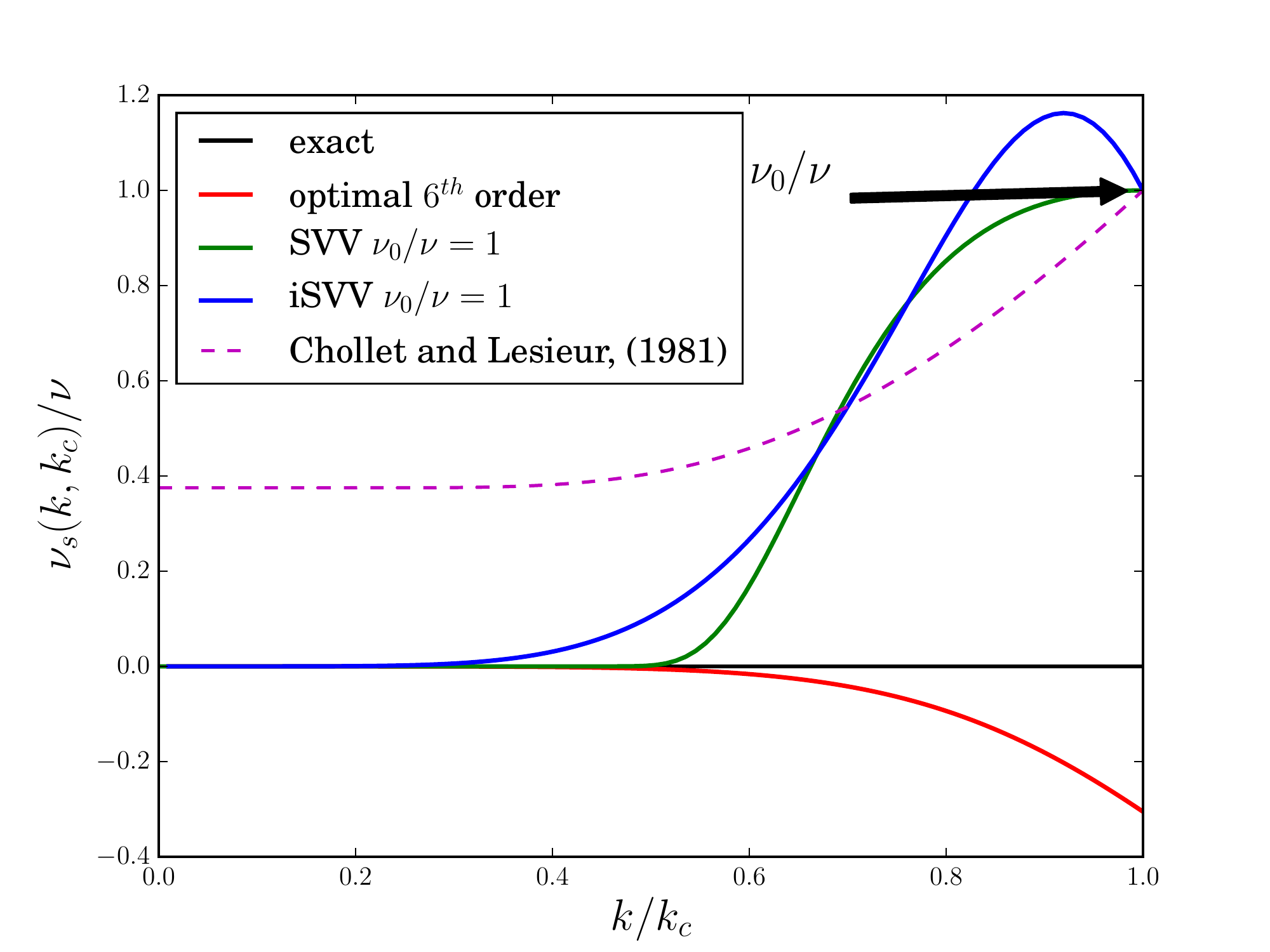}
\caption{}
\end{subfigure}
\begin{subfigure}[t]{.45\textwidth}
\centering
\includegraphics[width=\linewidth]{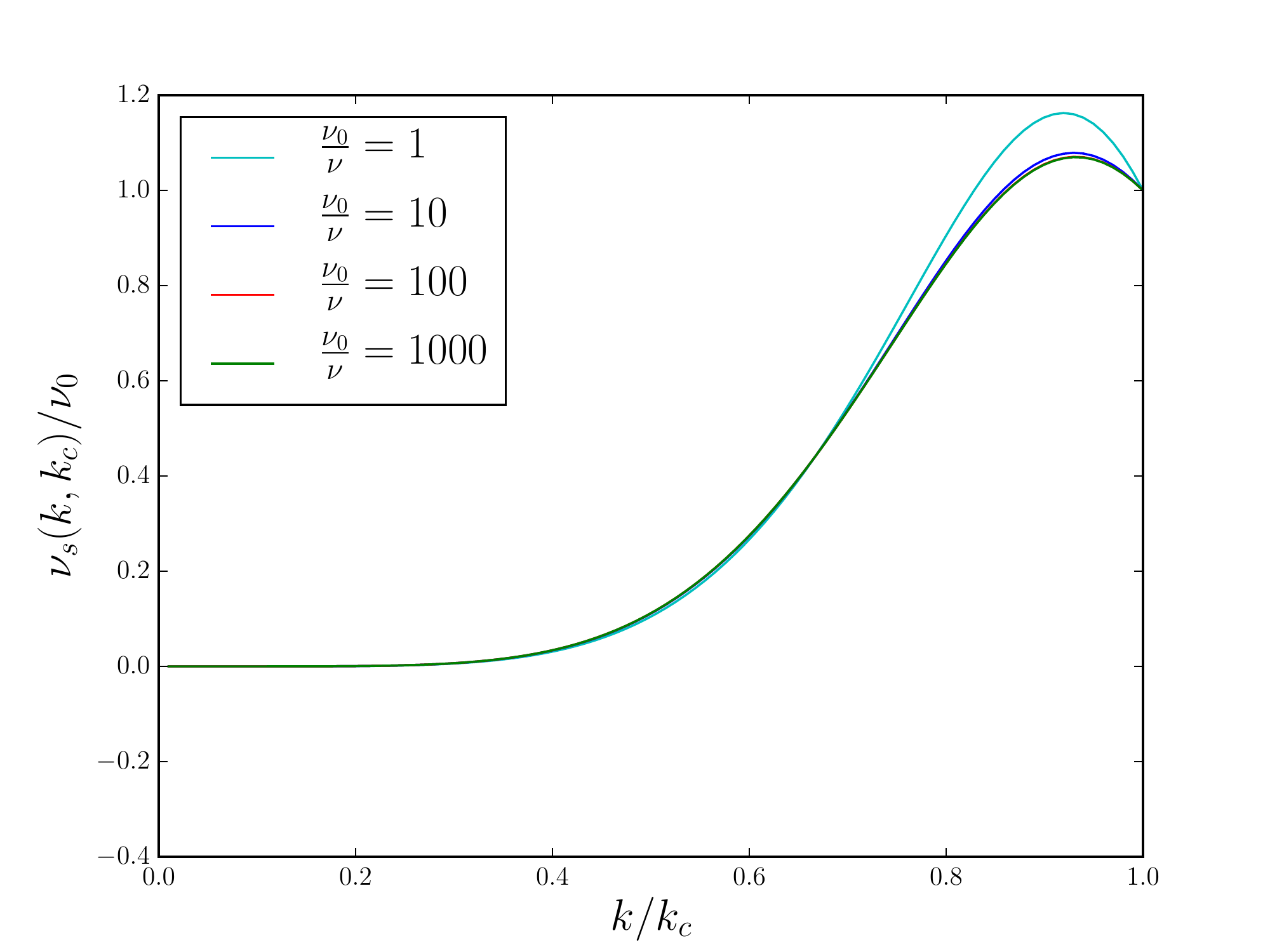}
\caption{}
\end{subfigure}
\caption{(a) The normalised associated spectral viscosity $\nu_s(k,k_c)/\nu$ of the exact, the ``optimal'' sixth--order accurate second derivative, as well as the implicit SVV (iSVV) operator plotted against the 
normalised wave number $k/k_c$. Also plotted is the explicit SVV kernel of \cite{Tadmor1989} and the spectral eddy viscosity of \cite{CholletLesieur1981}. 
(b) The implicit SVV operator spectral viscosity $\nu_s(k,k_c)/\nu$ scaled by the magnitude of SVV ($\nu_0/\nu$) for different values of $\nu_0/\nu$
plotted against the 
normalised wave number $k/k_c$. The figure confirms the linearity of the implicit SVV operator 
for values $\nu_0/\nu>$\num{10}.}
\label{fig:iSVV_kernel}
\end{figure*}
It can be easily observed from figure \ref{fig:iSVV_kernel}a that the implicit SVV kernel follows closely the explicit SVV kernel, thus justifying its name. 
The term `implicit' refers to the fact that the resulting operator is entirely due to the controlled numerical dissipation error. In addition, when the iSVV 
operator is used in the context of LES (not just to stabilise the solution), we may also refer to it as the iSVV--LES model. Another important observation which 
can be made for the same figure is that the "optimal" sixth--order scheme attains negative values near the cut-off, a fact which again proves its under-dissipative 
nature. Since we previously introduced the idea of a spectral eddy-viscosity model \citep{Kraichnan1976}, it is tempting to compare the physical-space constructed 
spectral viscosity $\nu_s(k)$ with existing spectral eddy-viscosity models such as that of \cite{CholletLesieur1981} which here we have scaled with the
kinematic viscosity $\nu$ to attain a maximum value of \num{1}:
\begin{equation}
\nu_s(k,k_c)=\nu K_0^{-3/2}[0.441+15.2\exp(-3.03k_c/k)],
\label{eq:CholletLesieurSEV}
\end{equation}
where $K_0=\,$\num{1.114}. The normalised \cite{CholletLesieur1981} spectral eddy viscosity is also plotted in figure 
\ref{fig:iSVV_kernel} and it is evident that the most significant difference is the low wavenumber ``plateau region'' of 
the spectral eddy viscosity which is missing from the SVV kernels and allows the spectral eddy viscosity to 
also affect larger flow scales. Next, looking at figure \ref{fig:iSVV_kernel}(b) it can be observed that the shape of the 
normalised spectral viscosities $\nu_s(k,k_c)/\nu_0$ is largely invariant to the SVV magnitude, a property which allows us to make 
the following approximation for the dissipation operator:
\begin{equation}
\mathcal{D}(u,a\nu_0/\nu)=a\mathcal{D}(u,\nu_0/\nu) \quad \text{ for any } \quad a\nu_0/\nu,\, \nu_0/\nu >10.
\label{eq:scaling_approx}
\end{equation}
This property, will become extremely useful in the formulation of the dynamic iSVV approach to follow, 
as the dissipation operator will be scaled by a local quantity. Nevertheless, the magnitude of the 
SVV $\nu_0/\nu$ depends on a number of parameters including the mesh resolution, Reynolds number ${Re}$,
as well as the accuracy of the underlying discretisation schemes. \cite{KaramanosKarniadakis2000} 
performed SVV--LES of a smooth wall turbulent channel flow using four variations of the SVV kernel parameters, including
its shape and magnitude, and compared with reference data. Unfortunately, from their simulations they were not able to identify 
a particular trend, although they suggested that an optimal value should exist and it should be sought through higher resolution 
simulations and a more meticulous comparison with the flow statistics. Later, \cite{KirbyKarniadakis2002} obtained
optimal values through comparison with reference data and higher--resolution computations. The most rigorous study to date for the
estimation of the optimal value for $\nu_0/\nu$  can be found in \cite{DairayEtAl2017} who used a Pao-like 
solution for the energy spectrum \citep{Pope2000} and were able to systematically calculate the optimal value, but unfortunately only
for the case of isotropic turbulence.

\subsection{The dynamic iSVV approach}\label{subsec:DSVV}
In the previous paragraph, we discussed the difficulties associated with selecting the right value for the magnitude of 
the spectral vanishing viscosity and the fact that most studies, except when isotropic turbulence is considered, have relied 
on choosing their values arbitrarily. However, what is also important to notice is that by considering a uniform 
distribution for the magnitude of the spectral viscosity within the computational domain, the SVV--LES approach
cannot take into account any localised flow features, and therefore dissipation is added irrespectively.  
\cite{KaramanosKarniadakis2000} recognised this defect of the static method in their concluding remarks and later \cite{KirbyKarniadakis2002} 
introduced and applied a dynamic SVV model, first to the inviscid Burgers equation and subsequently, to the 
compressible Navier--Stokes equations. The key idea underlying the dynamic SVV approach is to scale the 
amount of spectral viscosity by a localised flow diagnostic. \cite{KirbyKarniadakis2002} selected the magnitude of the local strain-rate tensor  
$\mathcal{S}=\sqrt{|2 S_{ij}S_{ij}|}$, where $S_{ij}=(\partial u_i /\partial x_j + \partial u_j 
/\partial x_i)/2$, and normalised the strain-rate tensor magnitude by a global scaling quantity 
$\mathcal{S}_{\infty}=\max\{\mathcal{S}\}$ taken over the entire domain. Here we have adopted the same scaling and we propose a new implementation 
of the implicit dynamic SVV model by invoking the linearity properties of the dissipation operator $\mathcal{D}$ (equation \eqref{eq:scaling_approx}), i.e.
\begin{equation}
\mathcal{D}\bigg(u_i,\frac{\mathcal{S}}{\mathcal{S}_{\infty}}\frac{\nu_0}{\nu}\bigg)=\frac{\mathcal{S}}{\mathcal{S}_{\infty}}\mathcal{D}\bigg(u_i,\frac{\nu_0}{\nu}\bigg).
\label{eq:dynFormulation}
\end{equation}
This approximation holds true only for large values of $\nu_0/\nu$, but allows us to implement the dynamic method without changing the numerical scheme. 
An alternative approach would need to interact with the numerical scheme by defining a local value for $\nu_0/\nu$. Another important feature of our proposed implementation is that the ratio ${\mathcal{S}}/\mathcal{S}_{\infty}$ can be calculated and monitored a priori so that additional 
constraints can be added, e.g. in the form of minimum or maximum values.  
\subsection{Actuator line model}\label{subsec:ALM}
For the turbine parametrisation, the actuator line model implementation of \cite{DeskosEtAl2017} 
and \cite{DeskosPiggott2017} is used. In particular, the rotor blades are parametrised into 
discrete blade elements and the normal and chordwise airload coefficients $C_n$, $C_c$  are 
computing using tabulated lift and drag coefficients of each blade element similar to 
\cite{SorensenShen2002} and \cite{TroldborgEtAl2009}. In addition the model accounts for dynamic stall 
effects, using a low--Mach number modification of the \cite{LeishmanBeddoes1989} model
proposed by \cite{ShengEtAl2008}, tip--loss correction effects \citep{ShenEtAl2005}, and a dynamic 
tower--shadow model similar to \cite{SarlakEtAl2015}. Finally, a standard Gaussian smoothing 
kernel approach is used to project the actuator line forces onto the fluid mesh, using a fixed smoothing
parameter $\epsilon=2.2 \Delta$, where $\Delta=(\Delta x \Delta y \Delta z)^{1/3}$.

\begin{figure*}[ht]
\centering
\begin{subfigure}[t]{0.4\textwidth}
\centering
\includegraphics[width=\linewidth]{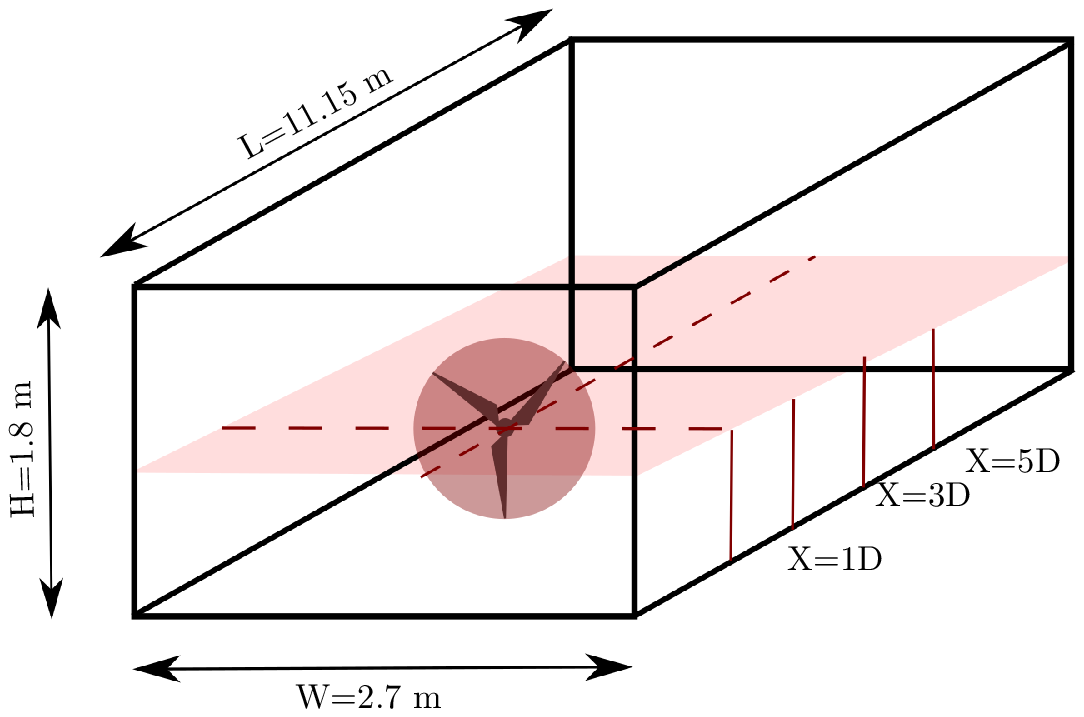}
\caption*{BT1}
\end{subfigure}
\begin{subfigure}[t]{0.43\textwidth}
\centering
\includegraphics[width=\linewidth]{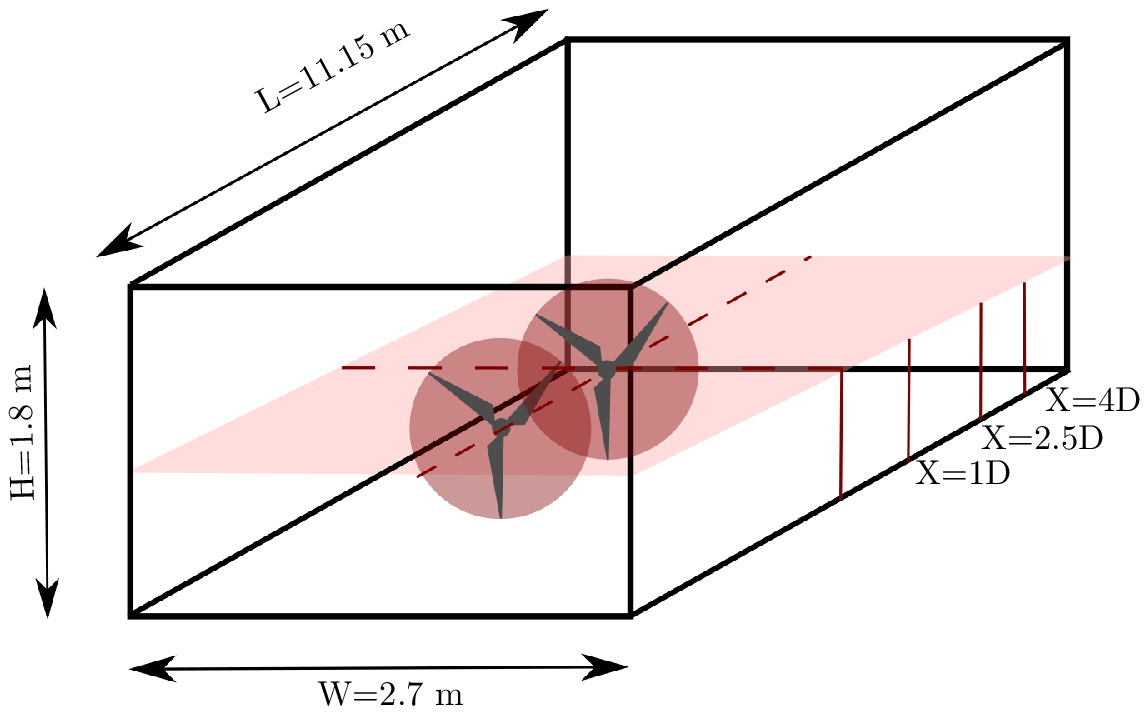}
\caption*{BT2}
\end{subfigure}
\caption{Schematic representation of the two blind tests (BT1 and BT2).}
\label{fig:BT_Schematic_Representation}
\end{figure*}

\section{Simulation set--up}\label{sec:SimSetup}
The numerical simulation set--ups are based on the wind tunnel experiments of \cite{KrogstadEriksen2013} and \cite{PierellaEtAl2014}. For brevity here we only
present information we feel is important to maintain clarity;
further details  on turbine blade 
characteristics or flow measurements techniques can be found in \cite{KrogstadEtAl2011} and \cite{KrogstadEriksen2013,PierellaEtAl2014}. The 
computational domain includes the full length of the wind tunnel which extends $L=\,$\SI{11.15}{\meter} in length, and has a width and height 
of $W=\,$\SI{2.7}{\meter} and $H=\,$\SI{1.8}{\meter} respectively. 
The measured incident velocity profile is found to be nearly uniform with $U_{\infty}=\,$\SI{10}{\meter \per \second} and turbulent intensity $\mathcal{I}=\,$\num{0.2}$\%$. For our simulations we introduce these
inflow conditions using ``planes'' generated by the isotropic synthetic fluctuation method of \cite{Davidson2007}. Standard 1D convective boundary conditions are used as outflow conditions in the streamwise direction and symmetric free-slip boundary conditions are implemented in the lateral directions. All simulations presented hereafter assume an effective rotor Reynolds number of $Re_D=\,$\num{100000}. For the turbine configurations we consider two cases: (1) a single turbine operating with a prescribed constant tip--speed ratio $\lambda=\,$ \num{6} which we will refer as the 
``blind test 1'' or BT1 hereafter, and (2) two similar turbines operating in line, with the second (rear) one being placed at a small distance 
(three diameters) behind the front one, which we will refer as the ``blind test 2'' or BT2. In the latter case the two turbines operate with 
prescribed rotational velocities which are different and in particular with tip speed ratios of $\lambda_1=\Omega_1 R/U_{\infty}=\,$\num{6} and $\lambda_2=\Omega_2 R/U_{\infty}=\,$\num{4}, where $\Omega_1$ and $\Omega_2$ are the rotor low speed angular velocities of the front and the rear turbines, respectively. 
A schematic representation of the two configurations is shown in figure 
\ref{fig:BT_Schematic_Representation}. The computational domain is discretised with \num{1281} $\times$ \num{241} $\times$ \num{241} mesh 
nodes, and a time step of $\Delta t\,$=\SI{0.0005}{\second} is used throughout. Each blade/actuator line is discretised with 70 
blade elements which ensures, together with the magnitude of $\Delta t$, stability and convergence for the actuator line/fluid solver coupling. The 
simulations were run for a total of \num{40000} time steps each and wake statistics were collected after a spin--up time of \SI{10}{\second} (\num{20000} 
time steps). Each of the simulations was performed with \num{576} computational cores and ran for approximately \SI{24}{\hour} on the Imperial College London cx2 supercomputer based on Intel Xeon E5-2680v3 (\num{2.5} GHz) processors. It should be noted that this wall clock time could easily be reduced by increasing the number of computational cores when running the 
simulations.
\section{Comparison to the wind tunnel data}\label{sec:Comparison}
\subsection{Static SVV wake solutions}\label{subsec:HeuristicStaticSVV}
Since a rigorous approach for the calculation of the required SVV magnitude $\nu_0/\nu$ does not yet exist for non--homogeneous turbulent flows, 
we have selected the following four values: \numlist{10;100;1000;2000} through an ``order-of-magnitude'' approach. 
\begin{figure*}[!ht]
\centering
\begin{subfigure}[t]{0.45\textwidth}
\centering
\includegraphics[width=\linewidth]{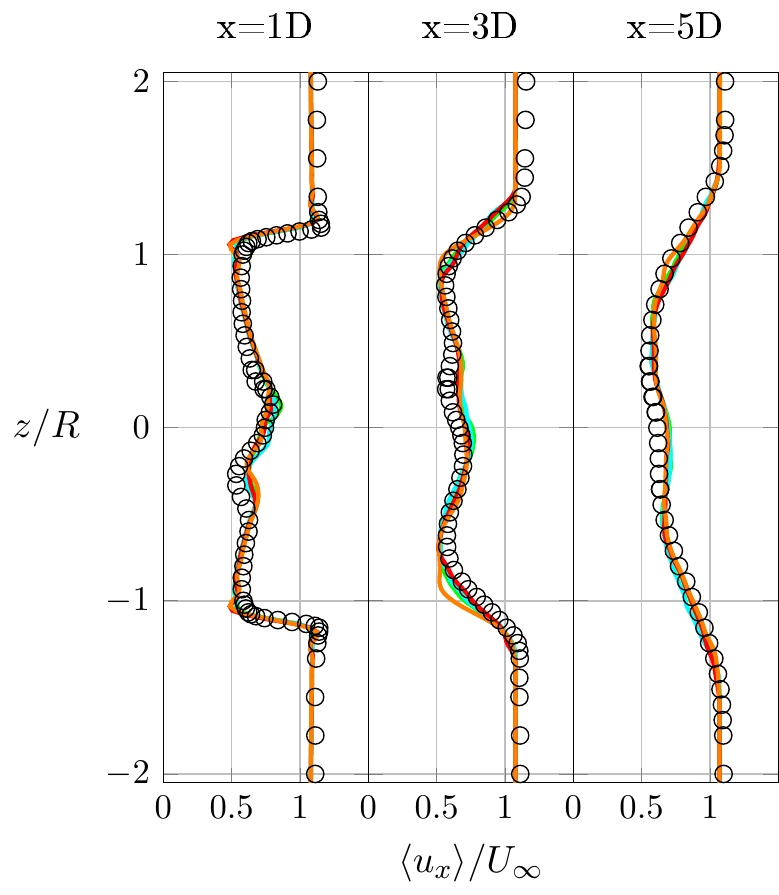}
\caption{}
\end{subfigure}
\begin{subfigure}[t]{0.45\textwidth}
\centering
\includegraphics[width=\linewidth]{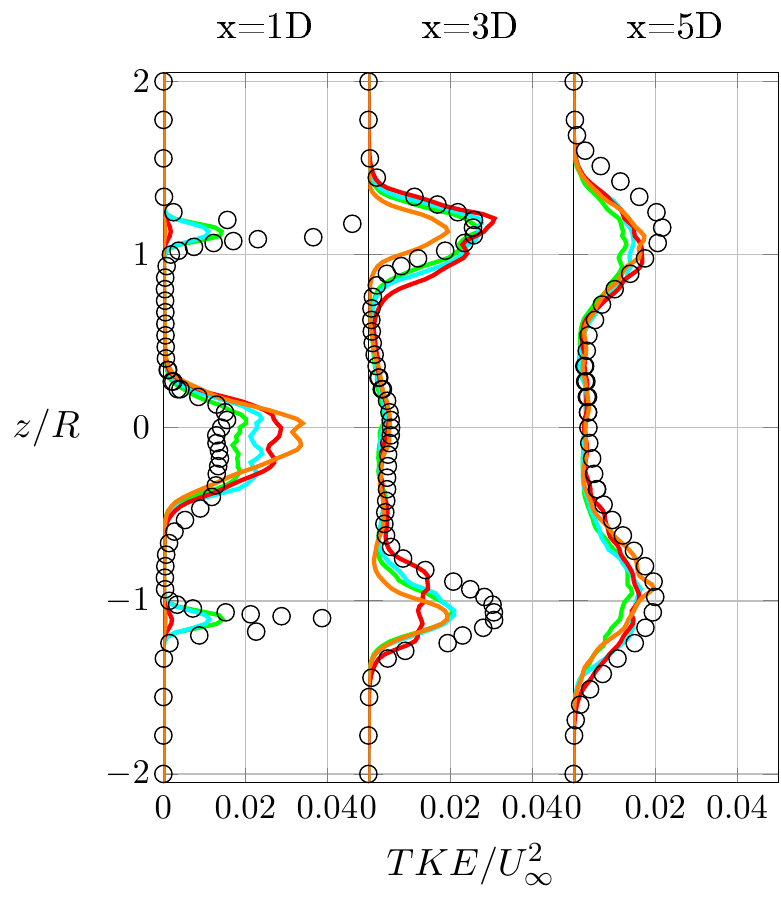}
\caption{}
\end{subfigure}
\\
\begin{subfigure}[ht]{0.45\textwidth}
\centering
\includegraphics[width=\linewidth]{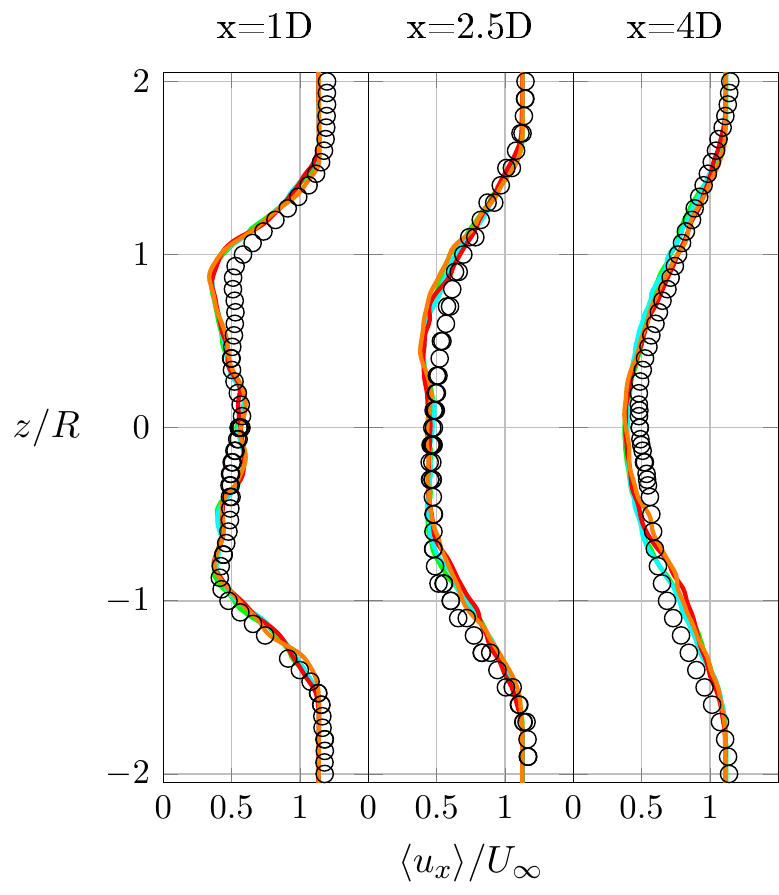}
\caption{}
\end{subfigure}
\begin{subfigure}[ht]{0.45\textwidth}
\centering
\includegraphics[width=\linewidth]{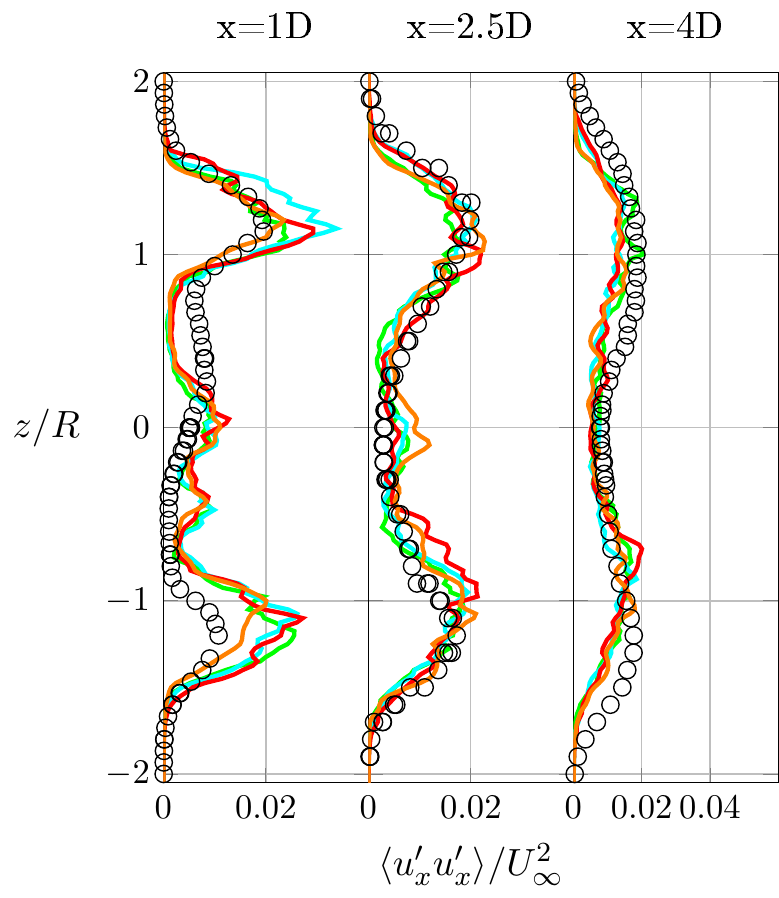}
\caption{}
\end{subfigure}
\caption{(a) Time--averaged horizontal profiles from BT1 for the streamwise velocity at $X/D=\,$\numlist{1;3;5} (b) Time--averaged turbulent kinetic energy (TKE) from BT1 at $X/D=\,$\numlist{1;3;5}. (c) Time--averaged horizontal profiles from BT2 for the streamwise velocity at $X/D=\,$\numlist{1;2.5;4} (d) Time--averaged turbulent kinetic energy (TKE) from BT1 at $X/D=\,$\numlist{1;2.5;4}. Solution for Static SVV $\nu_0/\nu=\,$\num{10} (\protect\greensolid), Static SVV $\nu_0/\nu=\,$\num{100} (\protect\cyansolid), Static SVV $\nu_0/\nu=\,$\num{1000} (\protect\redsolid), Static SVV $\nu_0/\nu=\,$\num{2000} (\protect\orangesolid) and experimental data from \cite{KrogstadEriksen2013} (\protect\Exp).}
\label{fig:Static}
\end{figure*}
The wake solutions obtained with the static implicit SVV method are compared against the reference wind tunnel data for both BT1 and BT2 in figure \ref{fig:Static}.
There we have plotted the horizontal time-averaged normalised streamwise velocity, $\langle u_x \rangle /U_{\infty}$, and the turbulence kinetic energy, 
TKE$/U_{\infty}^2$ at three locations downstream the rotor for BT1, and the horizontal time-averaged normalised streamwise velocity, $\langle u_x \rangle /U_{\infty}$, and streamwise Reynolds stresses $\langle u'_x u'_x \rangle /U_{\infty}^2$ at three locations downstream the rear turbine for BT2. 
All four wake solutions show good overall agreement with the wind tunnel measurements, particularly for the first-order statistics (mean velocity/velocity 
deficit). Many wake characteristics, including the near centreline asymmetry introduced by the tower wake, the width of the wake, as well as the magnitude of the 
velocity deficit are all very well predicted by the model, irrespective of the value of SVV. On the other hand, the magnitude of the SVV appears to have a more pronounced effect on the turbulent kinetic energy and streamwise Reynolds stresses profiles for BT1 and BT2, respectively. Starting with BT1, the low values 
$\nu_0/\nu=\,$\numlist{10;100} are shown to better capture the near wake ($x=1D$) TKE and in particular the magnitude of the TKE peak on the wake's outer thin 
shear layer, although the discrepancy between all wake solutions and the reference data remains high. This is related primarily to the ALM turbine parametrisation, 
and more particularly on the choice of a uniform smoothing value for $\epsilon$. It has been demonstrated that alternative formulations for the AL to mesh 
interpolation can mitigate these effects \citep{Martinez-TossasEtAl2017}. Nevertheless, it can be observed that an increase in $\nu_0/\nu$ will further smear out 
the two peaks. Paradoxically, an increase in the value of $\nu_0/\nu$ has an opposite effect on the the near centreline TKE. This is clearly an effect of the 
linearity of the SVV operator. An increased amount of dissipation will indiscriminately annihilate smaller turbulent scales even when they experience different 
levels of shear and therefore create a less distributed TKE profile. In other words, the computed sharp TKE distribution is a result of ``filtering'' the smaller 
scales and allowing only a large vortex to be shed from the tower. Again, the experimental data suggests that $\nu_0/\nu=\,$\numlist{10;100} are much more 
appropriate values. Looking at the downstream profiles of the TKE and more particularly at $x=3D$, the difference between the four simulations appears to be more 
distinguishable. This is a particular region in the spatial evolution of the wake where the most energetic structures are destabilised and break down into smaller
eddies. Inherently, an increase in the level of SVV dissipation may also affect the evolution of this process. The wake solution for $\nu_0/\nu=\,$\numlist{100;1000} 
yield the best match with the experimental data, which suggests that the optimal value should lie within these two values. On the other hand by applying a larger SVV value ($\nu_0/\nu=\,$\num{2000}) the solutions becomes over--dissipative, and lower levels of TKE are predicted. Finally, 
the wake solutions at $x=5D$ (onset of far wake field) appear to converge to the same solution. This result shows that variations in the SVV magnitude become less
important when the turbulent wake field becomes more isotropic. In BT2 the obtained wake statistics from the four different SVV parameters exhibit a very similar 
behaviour, although now the larger discrepancies appear at the profile immediately after the rear turbine ($x=1D$). 
This is due to the increased levels of TKE produced by the interaction of the upstream wake with the rear turbine, which manifests itself by first accelerating the 
breakup of the front turbine helical vortices, and subsequently shifting the merging/mixing zone upstream. It is worth noting that the solution for 
$\nu_0/\nu=\,$\num{2000}, although quantitatively closer to the reference data, is different from the other three solutions presented in figure \ref{fig:Static}d. 
From all these observations, we may conclude that an optimal value should lie between $\nu_0/\nu=\,$\num{100} and $\nu_0/\nu=\,$\num{1000}, although it cannot be 
established that this optimal value will out-perform all other values in terms of a rigorous quantitative metric. As we have previously mentioned, the linearity of the SVV approach will in general ignore the strength of the local flow features and will blindly add viscosity to the solution. Given the non--homogeneity of simulated turbine wake this might mean that for any selected value of the static SVV 
certain features of the wake, e.g. the merging zone, may be well captured whereas others may not. 
\subsection{Dynamic SVV wake solutions}\label{subsec:DynSVVResults}
Turning to the simulations using the dynamic SVV approach, we have chosen to run simulations only for the two higher values of SVV, 
$\nu_0/\nu=\,$\numlist{1000;2000}. This decision was driven by the current
formulation of the scaling factor $\mathcal{S}/\mathcal{S}_{\infty}$ which is bound
between \num{0} and \num{1}, and therefore the effective magnitude of the SVV
will be scaled down to $\bigg(\frac{\mathcal{S}}{\mathcal{S}_{\infty}}\frac{\nu_0}{\nu}<10\bigg)$ in places, which can potentially violate condition 
\eqref{eq:scaling_approx}. To guarantee that the local dynamic SVV magnitude does not go 
below \num{10} an extra condition has been imposed. This condition has a minimal impact
on our results, and the amount of minimum viscosity can be seen as a stabilisation for the solution.
\begin{figure*}[!ht]
\centering
\begin{subfigure}[ht]{0.45\textwidth}
\centering
\includegraphics[width=\linewidth]{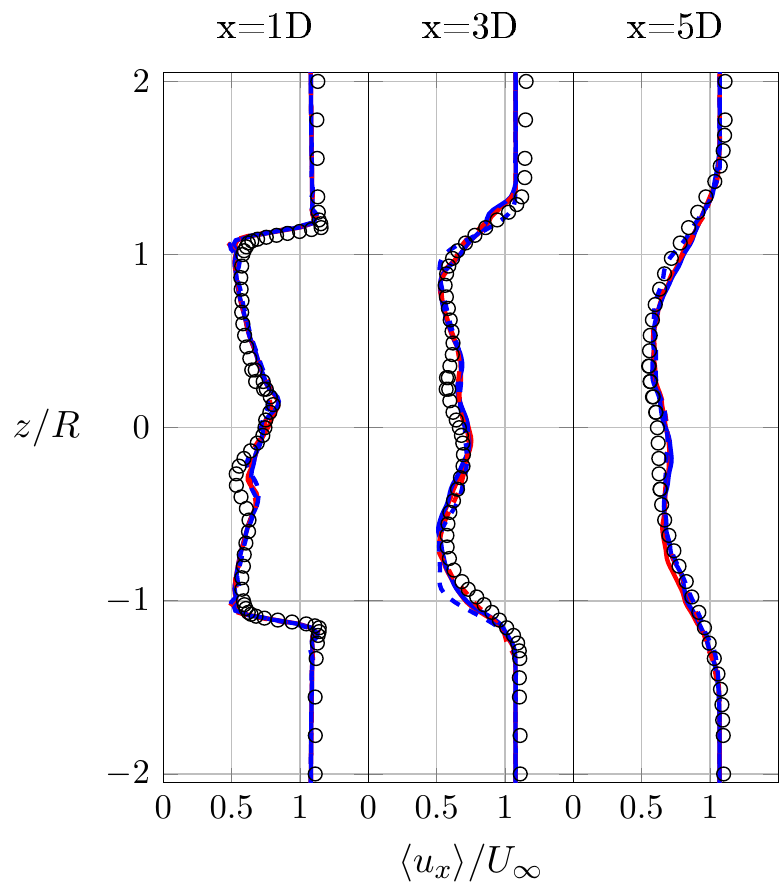}
\caption{}
\end{subfigure}
\begin{subfigure}[ht]{0.45\textwidth}
\centering
\includegraphics[width=\linewidth]{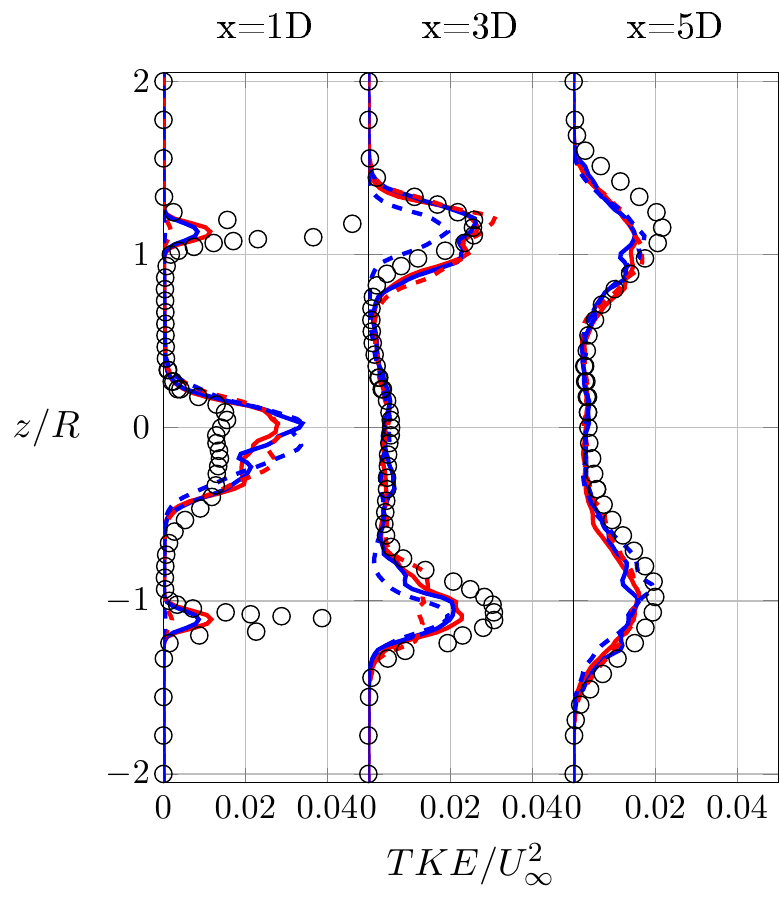}
\caption{}
\end{subfigure}
\\
\begin{subfigure}[ht]{0.45\textwidth}
\centering
\includegraphics[width=\linewidth]{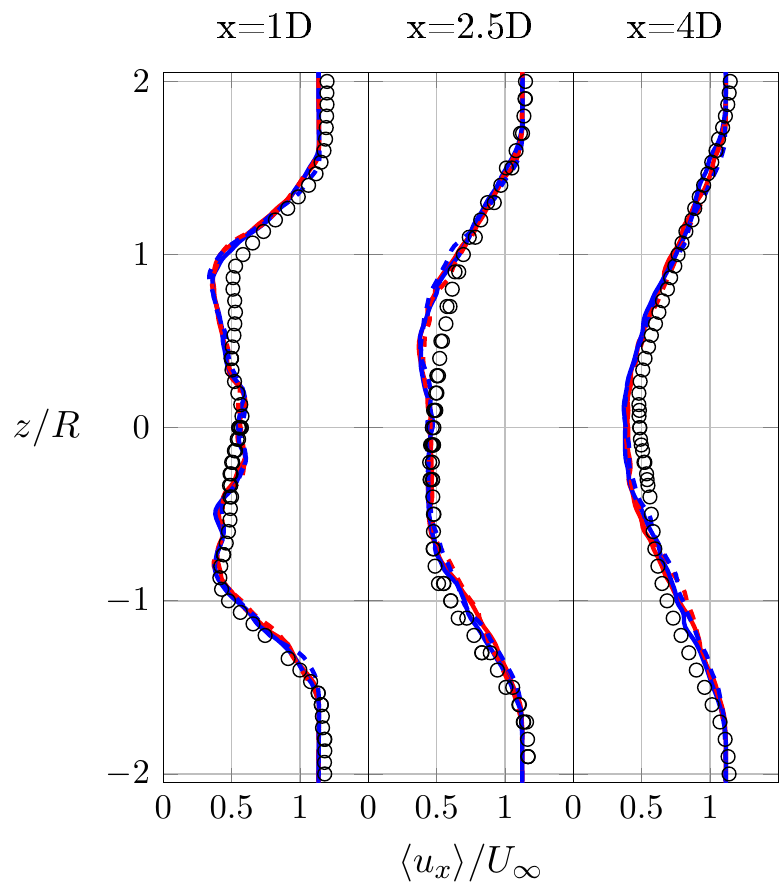}
\caption{}
\end{subfigure}
\begin{subfigure}[ht]{0.45\textwidth}
\centering
\includegraphics[width=\linewidth]{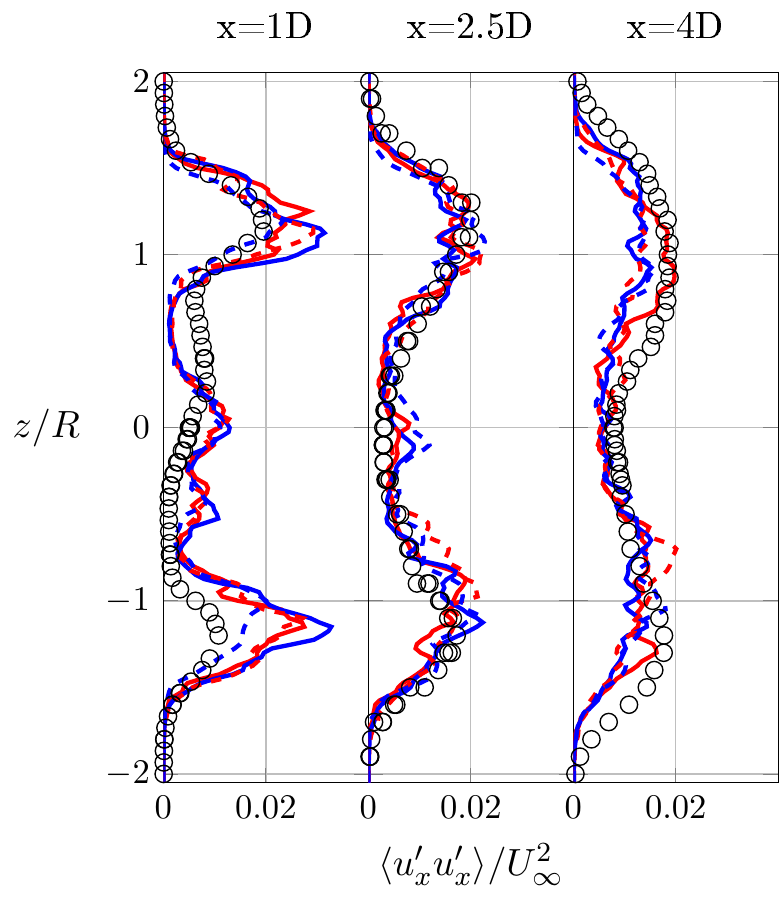}
\caption{}
\end{subfigure}
\caption{(a) Time--averaged horizontal profiles from BT1 for the streamwise velocity at $X/D=\,$\numlist{1;3;5} (b) Time--averaged turbulent kinetic energy (TKE) from BT1 at $X/D=\,$\numlist{1;3;5}. (c) Time--averaged horizontal profiles from BT2 for the streamwise velocity at $X/D=\,$\numlist{1;2.5;4} (d) Time--averaged turbulent kinetic energy (TKE) from BT1 at $X/D=\,$\numlist{1;2.5;4}. Solution for Static SVV $\nu_0/\nu=\,$\num{1000} (\protect\reddashed), Static SVV $\nu_0/\nu=\,$\num{2000} (\protect\bluedashed), Dynamic SVV $\nu_0/\nu=\,$\num{1000} (\protect\redsolid), Dynamic SVV $\nu_0/\nu=\,$\num{2000} (\protect\bluesolid) and experimental data from \cite{KrogstadEriksen2013} (\protect\Exp).}
\label{fig:BT1_StaticVsDyn}
\end{figure*}
Nevertheless, it is worth noting the quantitative differences between the static and
dynamic SVV approaches. An immediate observation is again that the two approaches agree
well as far as the mean streamwise profiles are concerned, although the dynamic method
appears to better capture the wake width, particularly in profile $x=3D$ of BT1. 
What is significant, however, is that with the dynamic method the discrepancies between 
the two SVV simulations have been significantly reduced. In particular, the TKE in the 
second profile ($x=3D$) of BT1 does not experience the sharp drop in the merging zone 
peaks as we change the SVV magnitude between $\nu_0/\nu=\,$\num{1000} and $\nu_0/\nu=\,$\num{2000} as happened before, and in addition the width of the shear layer 
(merging zone of the wake) is better captured. 
\subsection{Integral rotor quantities}\label{subsec:rotor}
Lastly, from the comparison with the wind tunnel data, we consider the rotor integrated quantities such as the power and thrust coefficients. These are not expected to differ significantly for either the static or the dynamic cases, or when varying SVV parameters. 
These is due to the fact that the SVV operator does not affect the large energetic scales. Nevertheless, the power and thrust coefficient are 
defined via,
\begin{subequations}
\begin{align}
C_P&=\frac{Q \Omega}{0.5 \rho A R U_{\infty}^3}, \\
C_T&=\frac{T}{0.5 \rho A U_{\infty}^2},
\end{align}
\end{subequations}
where $Q$ is the generated shaft torque, $\Omega$ the prescribed low speed rotational velocity and $A=\pi R^2$ the rotor's swept area.
We compute the time-averaged power and thrust coefficients after a spin-up time of \SI{20}{\second} which corresponds to at least 40 revolutions and the computed values are compared against the respective experimental values of \cite{KrogstadEriksen2013} and 
\cite{PierellaEtAl2014}. The final relative error estimates for all the simulations are shown in 
table \ref{table:IntegralQuantError}.
\begin{table*}[h]
\centering
\begin{center}
\captionof{table}{Error estimation for the integral rotor quantities for BT1 and BT2.}\label{table:IntegralQuantError}
\resizebox{\linewidth}{!}{\tiny
\begin{tabular}{ccccccc}
\hline
Test & SVV model & SVV param. & T1 $C_P$ error & T1 $C_T$ error & T2 $C_P$ error & T2 $C_T$ error\\
\hline
BT1 & Static    & \num{10}&   \num{-3.32}$\%$ & \num{8.89}$\%$ & - & - \\
BT1 & Static    & \num{100}&  \num{-3.28}$\%$ & \num{9.90}$\%$ & - & - \\
BT1 & Static    & \num{1000}& \num{-2.86}$\%$ & \num{9.90}$\%$ & - & - \\
BT1 & Static    & \num{2000}& \num{0.16}$\%$ & \num{10.27}$\%$ & - & - \\
BT1 & Dynamic   & \num{1000}& \num{-3.22}$\%$ & \num{8.93}$\%$ & - & - \\
BT1 & Dynamic   & \num{2000}& \num{-3.15}$\%$ & \num{8.95}$\%$ & - & - \\
BT2 & Static    & \num{10}&   \num{-6.06}$\%$ & \num{7.73}$\%$ & \num{-6.07}$\%$ & \num{-0.61}$\%$ \\
BT2 & Static    & \num{100}&  \num{-6.04}$\%$ & \num{7.74}$\%$ & \num{-6.21}$\%$ & \num{-0.60}$\%$ \\
BT2 & Static    & \num{1000}& \num{-5.74}$\%$ & \num{7.85}$\%$ & \num{-6.20}$\%$ & \num{-0.55}$\%$ \\
BT2 & Static    & \num{2000}& \num{-2.56}$\%$ & \num{9.11}$\%$ & \num{-7.55}$\%$ & \num{-1.23}$\%$ \\
BT2 & Dynamic   & \num{1000}& \num{-6.05}$\%$ & \num{7.73}$\%$ & \num{-6.01}$\%$ & \num{-0.27}$\%$ \\
BT2 & Dynamic   & \num{2000}& \num{-6.00}$\%$ & \num{7.75}$\%$ & \num{-7.07}$\%$ & \num{-0.96}$\%$ \\
\hline
\end{tabular}
}
\end{center}
\end{table*}
It is shown that all simulations exhibit similar levels for the relative errors for both the power and thrust coefficients. More 
specifically, in all simulations including both BT1 and BT2, the estimated power coefficients are over--predicted by approximately 
\num{-3}$\%$ for BT1 and \num{-5}$\%$ for the BT2 front turbine and \num{-6.5}$\%$ for the rear one. On the other hand,  $C_T$ is 
over--predicted for BT1 and the BT2 front turbine, while under--predicted for the rear BT2 one. Nevertheless, it is worth noting that 
for BT2 a bigger scatter is observed for the rear turbine quantities. This is not a surprising effect
as the second turbine is directly affected by the strength and turbulence structure of the upstream wake. We observe that the dynamic SVV for $\nu_0/\nu\,$=
\num{1000} provides the smallest error for both $C_P$ and $C_T$ for the rear turbine, while by increasing the dynamic SVV magnitude to $\nu_0/\nu\,$=\num{2000},
the error is increased by less that \num{1}$\%$ for both. Finally the values for BT1 and the front turbine of BT2, although they only experience inflow
isotropic turbulence,  also exhibit a better match with experimental values and less variability when the dynamic SVV method is used. This might be attributed to
the interaction of the near wake field with the tower wake which the rotor experiences as dynamic inflow.
 
\section{One--dimensional spectra and wake visualisations}\label{sec:AdditionalComparison}
One of the main advantages of LES is that it can capture the large coherent structures of turbulence with high accuracy. As we mentioned in the introduction, wind 
turbine wakes are well-known for their near-wake field helical vortices, as well as the particular mechanism which leads to them breaking up into smaller
eddies. Stability analysis of the near-wake helical vortices have been undertaken both in the context of LES 
\citep{IvanellEtAl2010,SarmastEtAl2014,SorensenEtAl2015} and analytical models \citep{Widnall1972,Okulov2004,OkulovSorensen2007}. In all these studies the main mechanism 
that leads to an effective breakup of the helical structures has been identified to be the merging and pairing of the neighbouring vortices caused by small 
perturbations present in the ambient flow. In an effort to highlight the ability of turbulence--resolving SVV simulations to capture the aforementioned mechanism as well as to demonstrate the advantages of the dynamic SVV, we present one--dimensional velocity spectra and wake visualisations for qualitative comparisons.
\subsection{One--dimensional spectra}\label{subsec:EnergySpectra}
\begin{figure*}[!ht]
\centering
\begin{subfigure}[t]{\textwidth}
\centering
\includegraphics[width=0.95\linewidth]{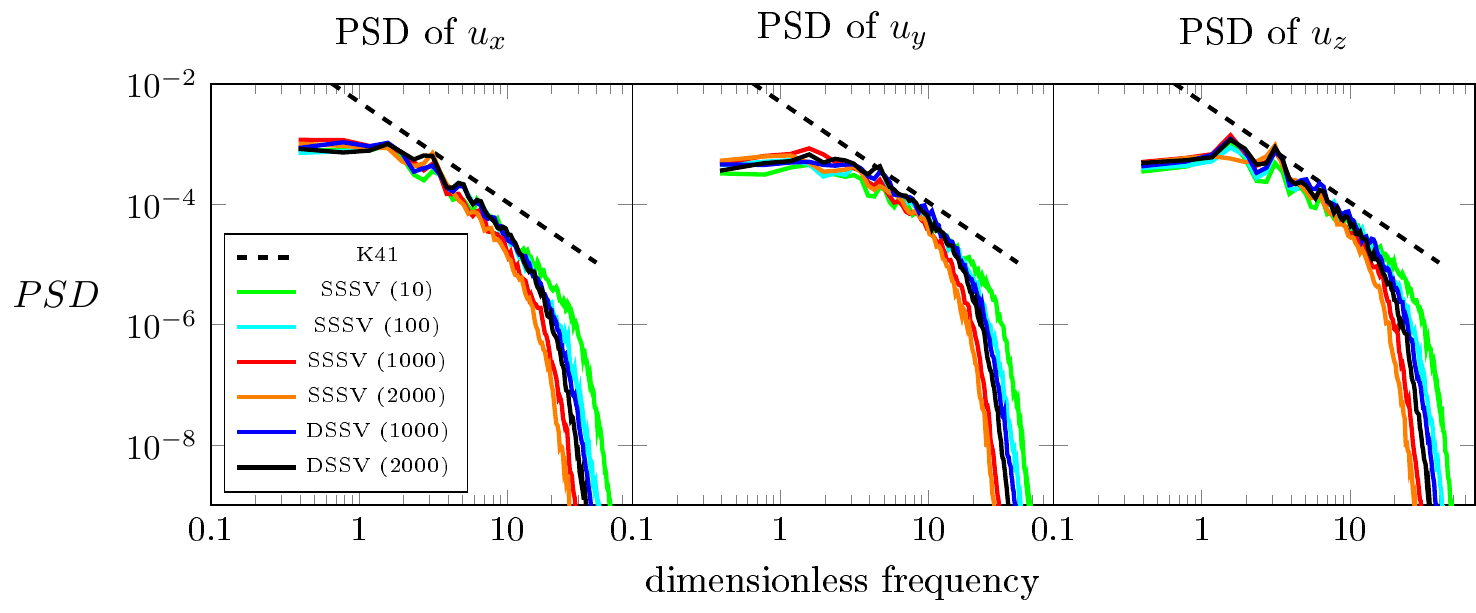}
\caption{}
\label{fig:BT1Spectra}
\end{subfigure}
\begin{subfigure}[t]{\textwidth}
\centering
\includegraphics[width=0.95\linewidth]{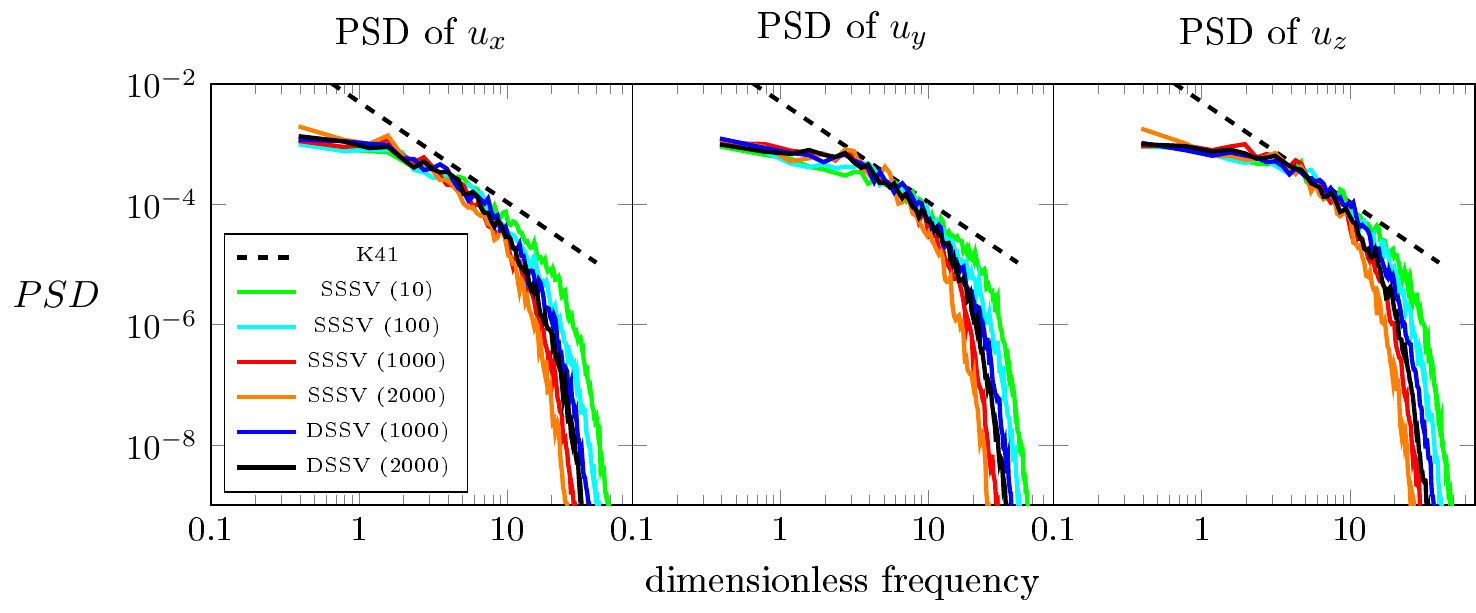}
\caption{}
\label{fig:BT2Spectra}
\end{subfigure}
\caption{Normalised power spectrum density (PSD) of $u_x$, $u_y$ and $u_z$ for all the assumed SVV magnitudes and the two approaches (static and dynamic).}
\end{figure*}

A better appreciation of the dissipation of the two methods (static and dynamic) as well as their respective selected parameters, can be sought through consideration of one--dimensional energy spectra. 
To obtain a meaningful estimate for the energy spectrum of the wake, we have deployed a total number of \num{30} ``probes'' (\num{15} along the centreline and \num{15} along the 
periphery of the wake, placed within half a meter distance each) downstream the turbines to sample the instantaneous velocities. Applying the frozen 
turbulence theory of Taylor \citep{Taylor1938,AntoniaEtAl1980}, we estimate the power spectrum density (PSD) for each component of the fluctuating velocity 
($u'_i=u_i-\langle u_i \rangle$) which are presented together with the characteristic inertial range $k^{-5/3}$ slope of \cite{Kolmogorov1941}, referred to as K41 hereafter in figures \ref{fig:BT1Spectra} 
and \ref{fig:BT2Spectra}
for BT1 and BT2, respectively.
The dimensionless power density spectra appear to follow similar trends irrespective of the selection of either the SVV magnitude for both BT1 and BT2. In particular,
in the lower frequency regime the different power spectral densities remain constant (large energetic scales) while subsequently all wake solutions follow the inertial sub--range $-5/3$ slope of 
the Kolmogorov spectrum \citep{Kolmogorov1941}. Perhaps the only region in which important differences can be observed is at the right--most regime. This is not surprising as the implicit
SVV model injects enhanced levels of dissipation according to the magnitude of SVV at the smaller scales and therefore it should affect the ``tail'' of the energy spectrum. Taking into account that the shape of the spectrum's tail is 
directly connected to the effective filter used in LES \citep{Sagaut2006}, the differences that are observed in all solutions confirm the ability of the implicit SVV--LES method to act as an explicit SGS
model. Moreover, the PSD obtained from the dynamic SVV solutions can better capture the energy distribution along the relevant scales, as it lies within the two curves $\nu_0/\nu=\,$\num{10} and 
$\nu_0/\nu=\,$\num{1000} in the right--most region. The exact effects of filtering may be obtained by computing the transfer function,
\begin{equation}
T_f(k)=\sqrt{\frac{E_{SVV}(k)}{E_{DNS}(k)}},
\end{equation}
where $E_{SVV}(k)$ is the wavenumber--dependent energy spectrum of the SVV simulations and $E_{DNS}(k)$ is the direct numerical simulation (DNS) respective one. Unfortunately, the high Reynolds number assumed 
for the simulations makes the computational cost of the respective DNS simulations prohibitively large.   
\subsection{Wake visualisation}
A common way to identify the coherent structures of a fluid flow is by using the so-called Q--criterion of \cite{HuntEtAl1988}, which describes the vortical 
structures via the positive second invariant of the velocity gradient. A similar technique was introduced by \cite{JeongHussain1995} with the so-called 
$\lambda_2$--criterion (not to be confused with the definition of the tip speed ratio used in the present work's nomenclature). To generate results comparable 
to previous studies \citep{SarlakEtAl2015,SorensenEtAl2015}, we make use of a non-standard vortex identification technique based on the magnitude of vorticity 
$|\omega|=\sqrt{\omega_x^2+\omega_y^2+\omega_z^2}$ with the same iso-surface $|\omega|=\,$\num{15} for all cases. Nevertheless, the assumed coherent structures 
allow us to observe the qualitative differences between the low dissipation and the high dissipation solutions. To appreciate the quality of the plotted coherent 
structures, it is worth re-iterating the mechanism of the near--wake destabilisation and evolution to the far--wake field. Instabilities start to appear at the end of the near--wake field and manifest themselves via intermediate--wave modes, which are multiples of the dominant tip vortex frequencies, and evolving along the tip vortices. \cite{Widnall1972} showed that such instabilities are due to the mutual--inductance of the tip vortices and that they have many common features with the 
vortex pair instability. Vortex pairing together with a subsequent vortex merging mechanism as was later demonstrated by 
\cite{SarmastEtAl2014} is responsible for the destabilisation of the near wake and its transition to the far wake field. 
We should emphasize at this point the importance of using a synthetic inlet method in triggering these modes and invoking the so--called 
``pairing--and--merging'' mechanism. Numerical experiments with grid-scale randomly generated inflow noise or uniform inflow profiles were found to be unable to 
yield the desired results.  
\begin{figure}[H]
\centering
\begin{subfigure}[t]{0.7\textwidth}
\centering
\includegraphics[width=0.9\linewidth]{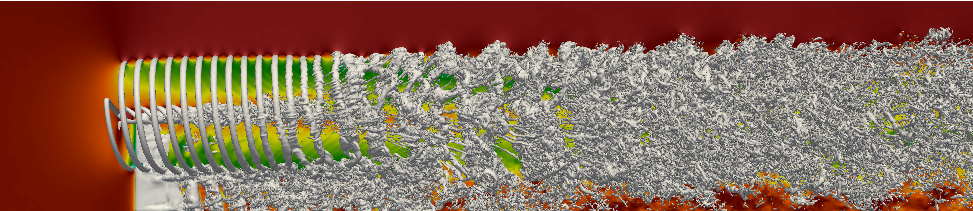}
\caption*{Static SVV, $\nu_0/\nu=10$}
\end{subfigure}
\begin{subfigure}[t]{0.7\textwidth}
\centering
\includegraphics[width=0.9\linewidth]{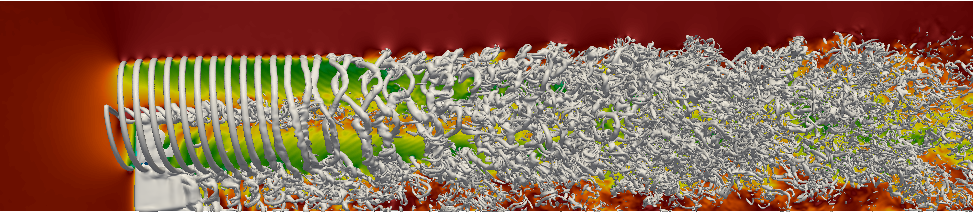}
\caption*{Static SVV, $\nu_0/\nu=100$}
\end{subfigure}
\begin{subfigure}[t]{0.7\textwidth}
\centering
\includegraphics[width=0.9\linewidth]{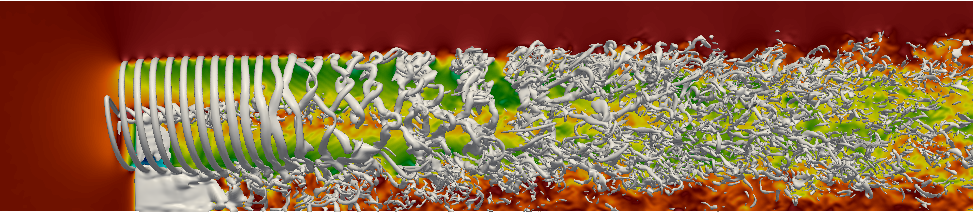}
\caption*{Static SVV, $\nu_0/\nu=1000$}
\end{subfigure}
\begin{subfigure}[t]{0.7\textwidth}
\centering
\includegraphics[width=0.9\linewidth]{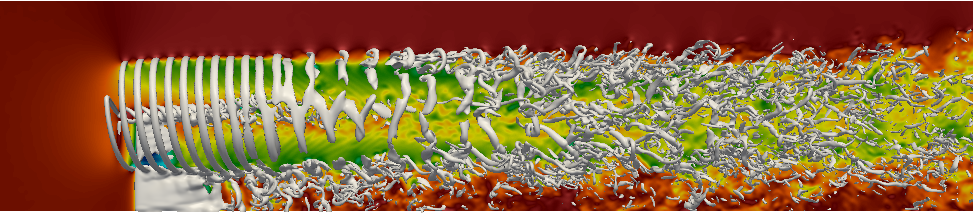}
\caption*{Static SVV, $\nu_0/\nu=2000$}
\end{subfigure}
\begin{subfigure}[t]{0.7\textwidth}
\centering
\includegraphics[width=0.9\linewidth]{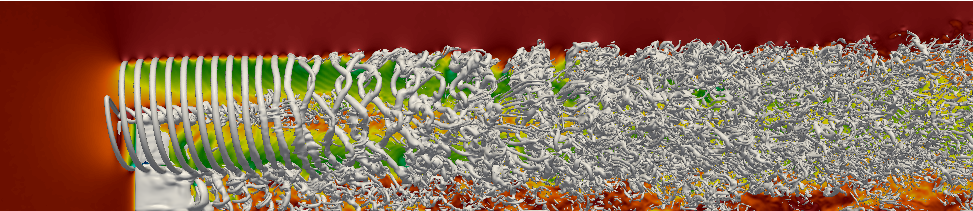}
\caption*{Dynamic SVV, $\nu_0/\nu=1000$}
\end{subfigure}
\begin{subfigure}[t]{0.7\textwidth}
\centering
\includegraphics[width=0.9\linewidth]{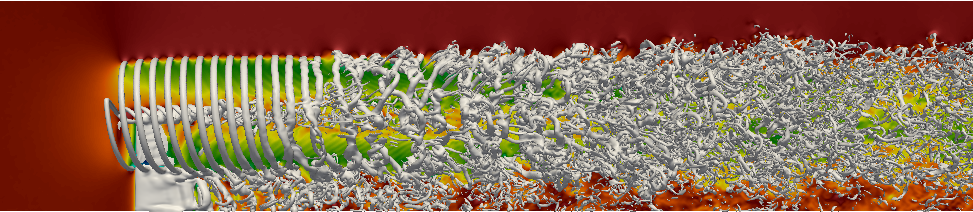}
\caption*{Dynamic SVV, $\nu_0/\nu=2000$}
\end{subfigure}
\caption{Wake visualisation for BT1: Contours of the instantaneous normalized enstrophy at $|\omega|=
\sqrt{\omega_x^2+\omega_y^2+\omega_z^2}=$\num{15} 
coloured with the magnitude of the streamwise velocity. Results are shown for static SVV with
$\nu_0/\nu=\,$\numlist{10;100;1000;2000} and dynamic SVV with $\nu_0/\nu=\,$\numlist{1000;2000}.  }
\label{fig:IsovorticityBT1}
\end{figure}
\begin{figure}[H]
\centering
\begin{subfigure}[t]{0.7\textwidth}
\centering
\includegraphics[width=0.9\linewidth]{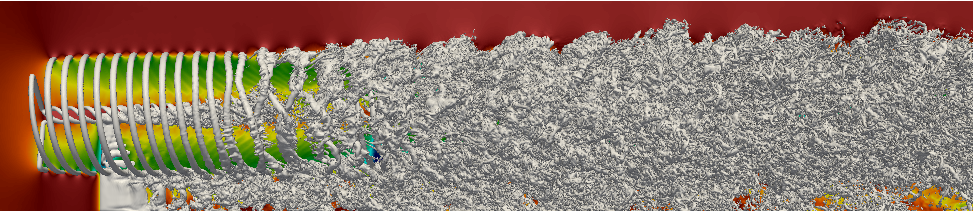}
\caption*{Static SVV, $\nu_0/\nu=10$}
\end{subfigure}
\begin{subfigure}[t]{0.7\textwidth}
\centering
\includegraphics[width=0.9\linewidth]{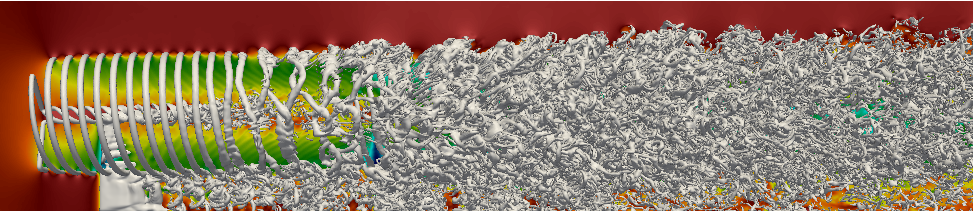}
\caption*{Static SVV, $\nu_0/\nu=100$}
\end{subfigure}
\begin{subfigure}[t]{0.7\textwidth}
\centering
\includegraphics[width=0.9\linewidth]{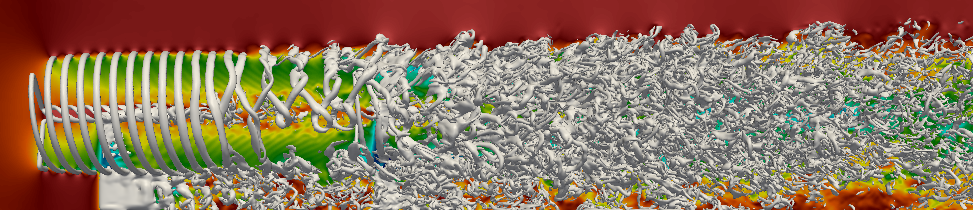}
\caption*{Static SVV, $\nu_0/\nu=1000$}
\end{subfigure}
\begin{subfigure}[t]{0.7\textwidth}
\centering
\includegraphics[width=0.9\linewidth]{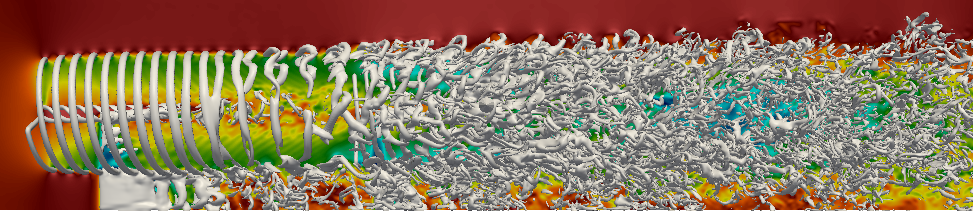}
\caption*{Static SVV, $\nu_0/\nu=2000$}
\end{subfigure}
\begin{subfigure}[t]{0.7\textwidth}
\centering
\includegraphics[width=0.9\linewidth]{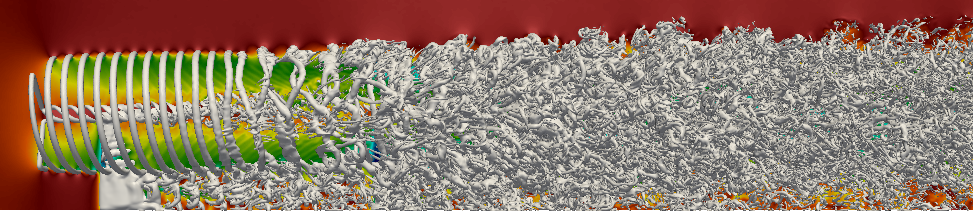}
\caption*{Dynamic SVV, $\nu_0/\nu=1000$}
\end{subfigure}
\begin{subfigure}[t]{0.7\textwidth}
\centering
\includegraphics[width=0.9\linewidth]{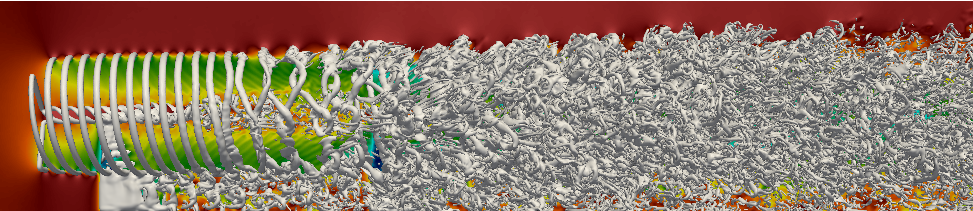}
\caption*{Dynamic SVV, $\nu_0/\nu=2000$}
\end{subfigure}
\caption{Wake visualisation for BT2: Caption similar to figure \ref{fig:IsovorticityBT1}.}
\label{fig:IsovorticityBT2}
\end{figure}
Looking at the obtained vorticity iso--contours plotted in figures \ref{fig:IsovorticityBT1} and \ref{fig:IsovorticityBT2}, we may observe that the above 
mentioned mechanism is visible only in the instantaneous coherent structures obtained with the static SVV approach for $\nu_0/\nu=\,$\num{100} and \num{1000} 
as well as the dynamic SVV approach for \num{1000} and \num{2000}. All other solutions do not capture these features satisfactory. For instance, the 
solutions obtained by using small amounts of dissipation (e.g. $\nu_0/\nu=\,$\num{10}) are found to exhibit short--wave instabilities, which most likely are a 
result of grid--scale spurious numerical oscillations. On the other hand, in SVV solutions with large amounts of numerical dissipation, the effective spectral viscosity would interact with the mutual--inductance modes/scales and therefore it will have an impact on the transition mechanism. This is 
confirmed by the wake visualisations of the static SVV for $\nu_0/\nu=\,$\num{2000}. There the pairing vortices have been merged due to the increased level of 
dissipation in the solution, and the expected instability modes have disappeared from the iso-contours. Apart from the transitional mechanism, the far wake field is 
also affected by increasing the magnitude of the SVV, as only the larger coherent structures remain in the solution. Thankfully, due to increasing levels of 
isotropy in the far wake field, an increase in the SVV magnitude will not affect any vital mechanism but instead it can be seen solely as an increase in the effective filtering as noted by \cite{DairayEtAl2017}. A ``golden ratio''
type of solution can be obtained by the dynamic SVV approach (e.g. DSVV with $\nu_0/\nu=\,$\num{1000}), as the local scaling of the dissipation acts more appropriately for both the near/transitional and far wake fields.

\section{Discussion and conclusions}\label{sec:Discussion}
We have investigated the ability of static and dynamic spectral vanishing viscosity (SVV) operators to act as a SGS model and predict the wake statistics behind 
a single wind turbine as well as two turbines operating in-line. This was motivated by previous studies in the field of wind turbine wakes 
\citep{MehtaEtAl2014,SarlakEtAl2015} which have pointed out that implicit LES formulations which rely on numerical 
dissipation for the SGS modelling can predict the wake equally well. In the present formulation we consider a higher-order compact finite--differences flow solver, 
the implicit SVV approach of \cite{DairayEtAl2017}, and a newly developed implicit Dynamic SVV approach to conduct LES of wind turbine wakes. The accuracy of the 
implicit SVV approach (static and dynamic) relies on the selection of the SVV magnitude and therefore a 
parametric analysis was required to understand its behaviour. To appreciate the sensitivity of the two approaches to the input parameter as well as determine an 
optimum value we perform comparisons against reference wind tunnel wake data. From this analysis we 
found that for the static SVV model an optimal value exists which provides a better match with the reference data, and this was found to lie between $\nu_0/\nu\,$=
\num{100} and \num{1000}. This optimum value was confirmed in both BT1 and BT2 test cases. On the other hand, the dynamic SVV showed a better match to reference 
data than static SVV, while a considerable variation of the $\nu_0/\nu$ parameter (from \num{1000} to \num{2000}) did not impact the quality of the results. 
A relative advantage of the implicit SVV models as compared to other more sophisticated SGS models is that they require no extra computational cost associated with 
the calculation of the implicit SVV term as their action is already incorporated in the numerical scheme. A minor increase in the computational cost is required by the dynamic SVV model due to the calculation of the local and global magnitudes of the shear rate tensor.  

More specifically, we followed a heuristic approach in order to conduct a parametric analysis for the magnitude of $\nu_0/\nu$ which is a priori unknown to us.
The values of $\nu_0/\nu$ were chosen with an ``order--of--magnitude'' based approach and the analysis showed that the optimal should lie within \numlist{100;1000}.
The linearity of the static SVV was shown to have two main drawbacks. First, as the same magnitude of dissipation was added to all the grid cut-off scales, 
irrespective of the local flow features (e.g. strong shear rate), even when using a nearly optimum value the produced solution created an ambiguous picture for the 
accuracy of the method. 
For instance, in the wake prediction for BT1 the strong near--wake shear layer was smeared out reducing the match with the experimental data while the downstream 
TKE predictions (near the merging zone region) agreed with them rather well. This means that while the value of \num{1000} is good for one region of the flow its 
performance is very poor for another. This effect was not observed in the study of \cite{DairayEtAl2017} and it is believed to be a feature of non-homogeneity 
as well as the interaction of the linear operator with the transitional part of the wake. The second drawback of the static method is related to the time-step
stability of the simulations, particularly because an explicit third--order Runge-Kutta method was used here. Inherently, in our simulations the viscous stability limit 
of \cite{Lele1992} constrains the time step in order to maintain stability in the numerical solution. When the magnitude of SVV exceeds a value of \num{1000} the 
initially assumed $\Delta t\,$=\SI{0.0005}{\second} did not satisfy the stability condition any more and therefore the time step had to be reduced by half. 
Surprisingly, the same condition was not required for the dynamic SVV approach, a finding which needs to be further examined in future studies. Nevertheless, apart 
from better time--step stability properties, the dynamic SVV addresses the problem of an indiscriminant dissipative action of the static SVV by 
scaling the local SVV magnitude with the magnitude of the local shear stress tensor. This property was shown to provide better results and rely less on the 
selection of the $\nu_0/\nu$ parameter. In fact the dynamic SVV approach has more similarities with the standard Smagorinsky model \citep{Smagorinsky1963}. Its 
dependence on the local flow features has been shown to provide better results both in a qualitative and quantitative sense. 

In summary, the present study has confirmed previous findings by showing that wind turbine wakes can be predicted with high accuracy if well--resolved simulations 
are considered, and that an implicit LES approach which depends on numerical dissipation can serve as an SGS model equally well. In the framework of SVV or SVV-like
models, the static method provides good results when an optimal magnitude value is used. However, a dynamic SVV approach would be more appropriate particularly when non-homogeneous turbulent flows are considered. Future studies will focus on improving the representation of the local scaling parameter so that the dynamic SVV method can be easily applied to homogeneous flows as well. We should also emphasize that the term ``dynamic'' can potentially create confusion as to the nature of the method. The term dynamic, is usually used to point out that the method behaves in an autonomous manner by adjusting all relevant parameters in order to achieve a goal (e.g. minimise energy dissipation). Here the term dynamic is used to signify that the value of SVV is not homogeneous but varies in space and in time with the magnitude of the shear-rate tensor. A truly dynamic method should be able to use flow diagnostics to determine the SVV magnitude, similar to the dynamic model of \cite{GermanoEtAl1991}. 

\section*{Acknowledgements}
The authors would like first and foremost to thank Professor Per-{\AA}ge Krogstad of NTNU for 
kindly providing an electronic version of the ``blind tests'' data. Funding from Imperial College London's Energy
Futures Lab and EPSRC (grant numbers EP/R007470/1 and EP/L000407/1) is also acknowledged. Finally, support for parallel 
computations was provided by Imperial College's Research Computing Service.

\bibliographystyle{elsarticle-harv}
\bibliography{Bibfile}

\begin{thebibliography}{72}
\expandafter\ifx\csname natexlab\endcsname\relax\def\natexlab#1{#1}\fi
\expandafter\ifx\csname url\endcsname\relax
  \def\url#1{\texttt{#1}}\fi
\expandafter\ifx\csname urlprefix\endcsname\relax\def\urlprefix{URL }\fi

\bibitem[{Ainslie(1988)}]{Ainslie1988}
Ainslie, J., 1988. Calculating the flowfield in the wake of wind turbines.
  Journal of Wind Engineering and Industrial Aerodynamics 27~(1), 213 -- 224.
\newline\urlprefix\url{http://www.sciencedirect.com/science/article/pii/0167610588900372}

\bibitem[{Antonia et~al.(1980)Antonia, Phan-Thien, and
  Chambers}]{AntoniaEtAl1980}
Antonia, R.~A., Phan-Thien, N., Chambers, A.~J., 9 1980. Taylor's hypothesis
  and the probability density functions of temporal velocity and temperature
  derivatives in a turbulent flow. Journal of Fluid Mechanics 100, 193--208.

\bibitem[{Calaf et~al.(2010)Calaf, Meneveau, and Meyers}]{CalafEtAl2010}
Calaf, M., Meneveau, C., Meyers, J., 2010. Large eddy simulation study of fully
  developed wind-turbine array boundary layers. Physics of Fluids 22~(1),
  015110.
\newline\urlprefix\url{http://dx.doi.org/10.1063/1.3291077}

\bibitem[{Chollet and Lesieur(1981)}]{CholletLesieur1981}
Chollet, J.-P., Lesieur, M., 1981. Parameterization of small scales of
  three-dimensional isotropic turbulence utilizing spectral closures. Journal
  of the Atmospheric Sciences 38~(12), 2747--2757.
\newline\urlprefix\url{https://doi.org/10.1175/1520-0469(1981)038<2747:POSSOT>2.0.CO;2}

\bibitem[{Churchfield et~al.(2012)Churchfield, Lee, Michalakes, and
  Moriarty}]{ChurchfieldEtAl2012}
Churchfield, M.~J., Lee, S., Michalakes, J., Moriarty, P.~J., 2012. A numerical
  study of the effects of atmospheric and wake turbulence on wind turbine
  dynamics. Journal of Turbulence 13, N14.
\newline\urlprefix\url{http://dx.doi.org/10.1080/14685248.2012.668191}

\bibitem[{Dairay et~al.(2014)Dairay, Fortuné, Lamballais, and
  Brizzi}]{DairayEtAl2014}
Dairay, T., Fortuné, V., Lamballais, E., Brizzi, L., 2014. {LES} of a
  turbulent jet impinging on a heated wall using high-order numerical schemes.
  International Journal of Heat and Fluid Flow 50, 177 -- 187.
\newline\urlprefix\url{http://www.sciencedirect.com/science/article/pii/S0142727X14000940}

\bibitem[{Dairay et~al.(2017)Dairay, Lamballais, Laizet, and
  Vassilicos}]{DairayEtAl2017}
Dairay, T., Lamballais, E., Laizet, S., Vassilicos, J.~C., 2017. Numerical
  dissipation vs. subgrid-scale modelling for large eddy simulation. Journal of
  Computational Physics 337, 252 -- 274.
\newline\urlprefix\url{http://www.sciencedirect.com/science/article/pii/S0021999117301298}

\bibitem[{Davidson(2007)}]{Davidson2007}
Davidson, L., 2007. Using isotropic synthetic fluctuations as inlet boundary
  conditions for unsteady simulations. In: Advances and Applications in Fluid
  Mechanics. Vol.~1. Pushpa Publishing House, pp. 1--35.

\bibitem[{Deskos et~al.(2017)Deskos, Abolghasemi, and Piggott}]{DeskosEtAl2017}
Deskos, G., Abolghasemi, M.~A., Piggott, M.~D., Aug 27--{S}ep 1 2017. Wake
  predictions from two turbine models using mesh-optimisation techniques. In:
  Lewis, A. (Ed.), {P}roceedings of the {T}welfth {E}uropean {W}ave and {T}idal
  {E}nergy {C}onference. {EWTEC}, University College Cork, Ireland, {ISSN}:
  2309-1983.

\bibitem[{Deskos and Piggott(2017)}]{DeskosPiggott2017}
Deskos, G., Piggott, M., 2017. Mesh-adaptive simulations of horizontal-axis
  turbine arrays using the actuator line method. ArXivSubmitted to Wind Energy
  Journal.

\bibitem[{Fischer and Mullen(2001)}]{FischerMullen2001}
Fischer, P., Mullen, J., 2001. Filter-based stabilization of spectral element
  methods. Comptes Rendus de l'Académie des Sciences - Series I - Mathematics
  332~(3), 265 -- 270.
\newline\urlprefix\url{http://www.sciencedirect.com/science/article/pii/S0764444200017638}

\bibitem[{Fischer et~al.(2008)Fischer, Lottes, and
  Kerkemeier}]{nek5000-web-page}
Fischer, P.~F., Lottes, J.~W., Kerkemeier, S.~G., 2008. {Nek5000} {W}eb page.
  Http://nek5000.mcs.anl.gov.

\bibitem[{Germano et~al.(1991)Germano, Piomelli, Moin, and
  Cabot}]{GermanoEtAl1991}
Germano, M., Piomelli, U., Moin, P., Cabot, W.~H., 1991. A dynamic
  subgrid--scale eddy viscosity model. Physics of Fluids A 3~(7), 1760--1765.
\newline\urlprefix\url{http://scitation.aip.org/content/aip/journal/pofa/3/7/10.1063/1.857955}

\bibitem[{Grinstein and Fureby(2004)}]{GrinsteinFureby2004}
Grinstein, F.~F., Fureby, C., Mar 2004. From canonical to complex flows: Recent
  progress on monotonically integrated les. Computing in Science Engineering
  6~(2), 36--49.

\bibitem[{Hunt et~al.(1988)Hunt, Wray, and Moin}]{HuntEtAl1988}
Hunt, J., Wray, A., Moin, P., 1988. Eddies, streams, and convergence zones in
  turbulent flows. In: Center for Turbulence Research.

\bibitem[{Ioannou and Laizet(2018)}]{IoannouLaizet2018}
Ioannou, V., Laizet, S., 2018. Numerical investigation of plasma-controlled
  turbulent jets for mixing enhancement. International Journal of Heat and
  Fluid Flow 70, 193 -- 205.
\newline\urlprefix\url{https://www.sciencedirect.com/science/article/pii/S0142727X17310342}

\bibitem[{Ivanell(2009)}]{Ivanell2009}
Ivanell, S., 2009. Numerical computations of wind turbine wakes. Ph.D. thesis,
  Royal Institute of Technology (KTH).

\bibitem[{Ivanell et~al.(2010)Ivanell, Mikkelsen, S{\o}rensen, and
  Henningson}]{IvanellEtAl2010}
Ivanell, S., Mikkelsen, R., S{\o}rensen, J.~N., Henningson, D., 2010. Stability
  analysis of the tip vortices of a wind turbine. Wind Energy 13~(8), 705--715.
\newline\urlprefix\url{http://dx.doi.org/10.1002/we.391}

\bibitem[{Jensen(1983)}]{Jensen1983}
Jensen, N., 1983. A note on wind generator interaction.

\bibitem[{Jeong and Hussain(1995)}]{JeongHussain1995}
Jeong, J., Hussain, F., 2 1995. On the identification of a vortex. Journal of
  Fluid Mechanics 285, 69--94.

\bibitem[{Jimenez et~al.(2007)Jimenez, Crespo, Migoya, and
  Garcia}]{JimenezEtAl2007}
Jimenez, A., Crespo, A., Migoya, E., Garcia, J., 2007. Advances in large-eddy
  simulation of a wind turbine wake. Journal of Physics: Conference Series
  75~(1), 012041.
\newline\urlprefix\url{http://stacks.iop.org/1742-6596/75/i=1/a=012041}

\bibitem[{Karamanos and Karniadakis(2000)}]{KaramanosKarniadakis2000}
Karamanos, G.-S., Karniadakis, G., 2000. A spectral vanishing viscosity method
  for large-eddy simulations. Journal of Computational Physics 163~(1), 22 --
  50.
\newline\urlprefix\url{http://www.sciencedirect.com/science/article/pii/S0021999100965525}

\bibitem[{Kirby and Karniadakis(2002)}]{KirbyKarniadakis2002}
Kirby, R., Karniadakis, G., 2002. Coarse resolution turbulence simulations with
  spectral vanishing viscosityâlarge-eddy simulations (svv-les). ASME. J.
  Fluids Eng. 124~(4).

\bibitem[{Kleusberg et~al.(2017)Kleusberg, Mikkelsen, Schlatter, Ivanell, and
  Henningson}]{KleusbergEtAl2017}
Kleusberg, E., Mikkelsen, R.~F., Schlatter, P., Ivanell, S., Henningson, D.~S.,
  2017. High-order numerical simulations of wind turbine wakes. Journal of
  Physics: Conference Series 854~(1), 012025.
\newline\urlprefix\url{http://stacks.iop.org/1742-6596/854/i=1/a=012025}

\bibitem[{Koal et~al.(2012)Koal, Stiller, and Blackburn}]{KoalEtAl2012}
Koal, K., Stiller, J., Blackburn, H., 2012. Adapting the spectral vanishing
  viscosity method for large-eddy simulations in cylindrical configurations.
  Journal of Computational Physics 231~(8), 3389 -- 3405.
\newline\urlprefix\url{http://www.sciencedirect.com/science/article/pii/S0021999112000356}

\bibitem[{Kolmogorov(1941)}]{Kolmogorov1941}
Kolmogorov, A., 1941. The local structure of turbulence in incompressible
  viscous fluid for very large reynolds numbers. Proc.Acad.Sci.URSS.

\bibitem[{Kraichnan(1976)}]{Kraichnan1976}
Kraichnan, R.~H., 1976. Eddy viscosity in two and three dimensions. Journal of
  the Atmospheric Sciences 33~(8), 1521--1536.
\newline\urlprefix\url{https://doi.org/10.1175/1520-0469(1976)033<1521:EVITAT>2.0.CO;2}

\bibitem[{Kravchenko and Moin(1997)}]{KravchenkoMoin1997}
Kravchenko, A., Moin, P., 1997. On the effect of numerical errors in large eddy
  simulations of turbulent flows. Journal of Computational Physics 131~(2), 310
  -- 322.
\newline\urlprefix\url{http://www.sciencedirect.com/science/article/pii/S0021999196955977}

\bibitem[{Krogstad et~al.(2011)Krogstad, Eriksen, and
  Melheim}]{KrogstadEtAl2011}
Krogstad, P.-A., Eriksen, P., Melheim, J., 2011. Blind test workshop;
  calculations for a model wind turbine. Tech. rep., Dept. Energy and Process
  Eng. NTNU.

\bibitem[{Krogstad and Eriksen(2013)}]{KrogstadEriksen2013}
Krogstad, P.-{\AA}., Eriksen, P.~E., 2013. ``blind test'' calculations of the
  performance and wake development for a model wind turbine. Renewable Energy
  50, 325 -- 333.
\newline\urlprefix\url{http://www.sciencedirect.com/science/article/pii/S0960148112003953}

\bibitem[{Laizet and Lamballais(2009)}]{LaizetLamballais2009}
Laizet, S., Lamballais, E., 2009. High-order compact schemes for incompressible
  flows: A simple and efficient method with quasi-spectral accuracy. Journal of
  Computational Physics 228~(16), 5989 -- 6015.
\newline\urlprefix\url{http://www.sciencedirect.com/science/article/pii/S0021999109002587}

\bibitem[{Laizet and Li(2011)}]{LaizetLi2011}
Laizet, S., Li, N., 2011. Incompact3d: A powerful tool to tackle turbulence
  problems with up to {O}(10$^5$) computational cores. International Journal
  for Numerical Methods in Fluids 67~(11), 1735--1757.
\newline\urlprefix\url{http://dx.doi.org/10.1002/fld.2480}

\bibitem[{Lamballais et~al.(2011)Lamballais, Fortuné, and
  Laizet}]{LamballaisEtAl2011}
Lamballais, E., Fortuné, V., Laizet, S., 2011. Straightforward high-order
  numerical dissipation via the viscous term for direct and large eddy
  simulation. Journal of Computational Physics 230~(9), 3270 -- 3275.
\newline\urlprefix\url{http://www.sciencedirect.com/science/article/pii/S0021999111000659}

\bibitem[{Leishman and Beddoes(1989)}]{LeishmanBeddoes1989}
Leishman, J.~G., Beddoes, T.~S., 1989. A semi-empirical model for dynamic
  stall. Journal of the American Helicopter Society 34.

\bibitem[{Lele(1992)}]{Lele1992}
Lele, S.~K., 1992. Compact finite difference schemes with spectral-like
  resolution. Journal of Computational Physics 103~(1), 16 -- 42.
\newline\urlprefix\url{http://www.sciencedirect.com/science/article/pii/002199919290324R}

\bibitem[{Lombard et~al.(2016)Lombard, Moxey, Sherwin, Hoessler, Dhandapani,
  and Taylor}]{LombardEtAl2016}
Lombard, J.-E.~W., Moxey, D., Sherwin, S.~J., Hoessler, J. F.~A., Dhandapani,
  S., Taylor, M.~J., 2016. Implicit large-eddy simulation of a wingtip vortex.
  AIAA Journal.

\bibitem[{Lu and Port{\'e}-Agel(2011)}]{LuPorte-Agel2011}
Lu, H., Port{\'e}-Agel, F., 2011. Large-eddy simulation of a very large wind
  farm in a stable atmospheric boundary layer. Physics of Fluids 23~(6).
\newline\urlprefix\url{http://scitation.aip.org/content/aip/journal/pof2/23/6/10.1063/1.3589857}

\bibitem[{Mart{\'i}nez-Tossas et~al.(2017)Mart{\'i}nez-Tossas, Churchfield, and
  Meneveau}]{Martinez-TossasEtAl2017}
Mart{\'i}nez-Tossas, L.~A., Churchfield, M.~J., Meneveau, C., 2017. Optimal
  smoothing length scale for actuator line models of wind turbine blades based
  on {G}aussian body force distribution. Wind Energy, n/a--n/aWe.2081.
\newline\urlprefix\url{http://dx.doi.org/10.1002/we.2081}

\bibitem[{Mehta et~al.(2014)Mehta, van Zuijlen, Koren, Holierhoek, and
  Bijl}]{MehtaEtAl2014}
Mehta, D., van Zuijlen, A., Koren, B., Holierhoek, J., Bijl, H., 2014. Large
  eddy simulation of wind farm aerodynamics: A review. Journal of Wind
  Engineering and Industrial Aerodynamics 133, 1 -- 17.
\newline\urlprefix\url{http://www.sciencedirect.com/science/article/pii/S0167610514001391}

\bibitem[{Meneveau and Katz(2000)}]{MeneveauKatz2000}
Meneveau, C., Katz, J., 2000. Scale-invariance and turbulence models for
  large-eddy simulation. Annual Review of Fluid Mechanics 32~(1), 1--32.
\newline\urlprefix\url{https://doi.org/10.1146/annurev.fluid.32.1.1}

\bibitem[{Mengaldo et~al.(2017)Mengaldo, Moura, Giralda, Peir{\'o}, and
  Sherwin}]{MengaldoEtAl2017}
Mengaldo, G., Moura, R., Giralda, B., Peir{\'o}, J., Sherwin, S., 2017. Spatial
  {E}igensolution analysis of discontinuous {G}alerkin schemes with practical
  insights for under-resolved computations and implicit {LES}. Computers \&
  Fluids.
\newline\urlprefix\url{http://www.sciencedirect.com/science/article/pii/S0045793017303511}

\bibitem[{Nilsson et~al.(2015)Nilsson, Ivanell, Hansen, Mikkelsen, S{\o}rensen,
  Breton, and Henningson}]{NilssonEtAl2015}
Nilsson, K., Ivanell, S., Hansen, K.~S., Mikkelsen, R., S{\o}rensen, J.~N.,
  Breton, S.-P., Henningson, D., 2015. Large-eddy simulations of the lillgrund
  wind farm. Wind Energy 18~(3), 449--467.
\newline\urlprefix\url{http://dx.doi.org/10.1002/we.1707}

\bibitem[{Okulov(2004)}]{Okulov2004}
Okulov, V.~L., 12 2004. On the stability of multiple helical vortices. Journal
  of Fluid Mechanics 521, 319--342.

\bibitem[{Okulov and S{\o}rensen(2007)}]{OkulovSorensen2007}
Okulov, V.~L., S{\o}rensen, J.~N., 4 2007. Stability of helical tip vortices in
  a rotor far wake. Journal of Fluid Mechanics 576, 1--25.

\bibitem[{Orszag(1970)}]{Orszag1970}
Orszag, S.~A., 1970. Transform method for the calculation of vector-coupled
  sums: Application to the spectral form of the vorticity equation. Journal of
  the Atmospheric Sciences 27~(6), 890--895.
\newline\urlprefix\url{https://doi.org/10.1175/1520-0469(1970)027<0890:TMFTCO>2.0.CO;2}

\bibitem[{Pasquetti(2005)}]{Pasquetti2005}
Pasquetti, R., 2005. Spectral vanishing viscosity method for les: sensitivity
  to the svv control parameters. Journal of Turbulence 6, N12.
\newline\urlprefix\url{https://doi.org/10.1080/14685240500125476}

\bibitem[{Pasquetti(2006)}]{Pasquetti2006}
Pasquetti, R., Jun 2006. Spectral vanishing viscosity method for large-eddy
  simulation of turbulent flows. Journal of Scientific Computing 27~(1),
  365--375.
\newline\urlprefix\url{https://doi.org/10.1007/s10915-005-9029-9}

\bibitem[{Pasquetti et~al.(2008)Pasquetti, S{\'e}verac, Serre, Bontoux, and
  Sch{\"a}fer}]{PasquettiEtAl2008}
Pasquetti, R., S{\'e}verac, E., Serre, E., Bontoux, P., Sch{\"a}fer, M., May
  2008. From stratified wakes to rotor--stator flows by an {SVV--LES} method.
  Theoretical and Computational Fluid Dynamics 22~(3), 261--273.
\newline\urlprefix\url{https://doi.org/10.1007/s00162-007-0070-1}

\bibitem[{Peet et~al.(2013)Peet, Fischer, Conzelmann, and
  Kotamarthi}]{PeetEtAl2013}
Peet, Y., Fischer, P., Conzelmann, G., Kotamarthi, V., 2013. Actuator line
  aerodynamics model with spectral elements. AIAA Journal.

\bibitem[{Pierella et~al.(2014)Pierella, Krogstad, and
  S{\ae}tran}]{PierellaEtAl2014}
Pierella, F., Krogstad, P.-{\AA}., S{\ae}tran, L., 2014. Blind test 2
  calculations for two in-line model wind turbines where the downstream turbine
  operates at various rotational speeds. Renewable Energy 70, 62 -- 77, special
  issue on aerodynamics of offshore wind energy systems and wakes.
\newline\urlprefix\url{http://www.sciencedirect.com/science/article/pii/S0960148114001815}

\bibitem[{Pope(2000)}]{Pope2000}
Pope, S., 2000. Turbulent Flows. Cambridge University Press.
\newline\urlprefix\url{https://books.google.com/books?id=HZsTw9SMx-0C}

\bibitem[{Rai and Moin(1991)}]{RaiMoin1991}
Rai, M.~M., Moin, P., 1991. Direct simulations of turbulent flow using
  finite-difference schemes. Journal of Computational Physics 96~(1), 15 -- 53.
\newline\urlprefix\url{http://www.sciencedirect.com/science/article/pii/002199919190264L}

\bibitem[{Sagaut(2006)}]{Sagaut2006}
Sagaut, P., 2006. Large Eddy Simulation for Incompressible Flows: An
  Introduction. Scientific Computation. Springer.
\newline\urlprefix\url{https://books.google.com/books?id=ODYiH6RNyoQC}

\bibitem[{Sanderse et~al.(2011)Sanderse, van~der Pijl, and
  Koren}]{SanderseEtAl2011}
Sanderse, B., van~der Pijl, S., Koren, B., 2011. Review of computational fluid
  dynamics for wind turbine wake aerodynamics. Wind Energy 14~(7), 799--819.
\newline\urlprefix\url{http://dx.doi.org/10.1002/we.458}

\bibitem[{Sarlak(2014)}]{Sarlak2014}
Sarlak, H., 2014. Large eddy simulation of turbulent flows in wind energy.
  Ph.D. thesis, Technical University of Denmark.

\bibitem[{Sarlak et~al.(2015)Sarlak, Meneveau, and
  S{\o}rensen}]{SarlakEtAl2015}
Sarlak, H., Meneveau, C., S{\o}rensen, J., 2015. Role of subgrid-scale modeling
  in large eddy simulation of wind turbine wake interactions. Renewable Energy
  77, 386 -- 399.
\newline\urlprefix\url{http://www.sciencedirect.com/science/article/pii/S0960148114008635}

\bibitem[{Sarmast et~al.(2014)Sarmast, Dadfar, Mikkelsen, Schlatter, Ivanell,
  S{\o}rensen, and Henningson}]{SarmastEtAl2014}
Sarmast, S., Dadfar, R., Mikkelsen, R.~F., Schlatter, P., Ivanell, S.,
  S{\o}rensen, J., Henningson, D., 2014. Mutual inductance instability of the
  tip vortices behind a wind turbine. Journal of Fluid Mechanics 755,
  705â731.

\bibitem[{Severac and Serre(2007)}]{SeveracSerre2007}
Severac, E., Serre, E., 2007. A spectral vanishing viscosity for the {LES} of
  turbulent flows within rotating cavities. Journal of Computational Physics
  226~(2), 1234 -- 1255.
\newline\urlprefix\url{http://www.sciencedirect.com/science/article/pii/S0021999107002318}

\bibitem[{Shen et~al.(2005)Shen, Mikkelsen, S{\o}rensen, and
  Bak}]{ShenEtAl2005}
Shen, W.~Z., Mikkelsen, R., S{\o}rensen, J.~N., Bak, C., 2005. Tip loss
  corrections for wind turbine computations. Wind Energy 8~(4), 457--475.
\newline\urlprefix\url{http://dx.doi.org/10.1002/we.153}

\bibitem[{Sheng et~al.(2008)Sheng, Galbraith, and Coton}]{ShengEtAl2008}
Sheng, W., Galbraith, R.~A., Coton, F.~N., 2008. A modified dynamic stall model
  for low mach numbers. Journal of Solar Energy Engineering.

\bibitem[{Smagorinsky(1963)}]{Smagorinsky1963}
Smagorinsky, J., 1963. General circulation experiments with the primitive
  equations. Monthly Weather Review 91~(3), 99--164.
\newline\urlprefix\url{http://dx.doi.org/10.1175/1520-0493(1963)091<0099:GCEWTP>2.3.CO;2}

\bibitem[{S{\o}rensen and Shen(2002)}]{SorensenShen2002}
S{\o}rensen, J., Shen, W.~Z., May 2002. Numerical modeling of wind turbine
  wakes. Journal of Fluids Engineering 124~(2), 393--399, 2002.

\bibitem[{S{\o}rensen et~al.(2015)S{\o}rensen, Mikkelsen, Henningson, Ivanell,
  Sarmast, and Andersen}]{SorensenEtAl2015}
S{\o}rensen, J.~N., Mikkelsen, R.~F., Henningson, D.~S., Ivanell, S., Sarmast,
  S., Andersen, S.~J., 2015. Simulation of wind turbine wakes using the
  actuator line technique. Philosophical Transactions of the Royal Society of
  London A: Mathematical, Physical and Engineering Sciences 373~(2035).
\newline\urlprefix\url{http://rsta.royalsocietypublishing.org/content/373/2035/20140071}

\bibitem[{Tadmor(1989)}]{Tadmor1989}
Tadmor, E., 1989. Convergence of spectral methods for nonlinear conservation
  laws. SIAM Journal on Numerical Analysis 26~(1), 30--44.
\newline\urlprefix\url{https://doi.org/10.1137/0726003}

\bibitem[{Taylor(1938)}]{Taylor1938}
Taylor, G.~I., 1938. The spectrum of turbulence. Proceedings of the Royal
  Society of London A: Mathematical, Physical and Engineering Sciences
  164~(919), 476--490.
\newline\urlprefix\url{http://rspa.royalsocietypublishing.org/content/164/919/476}

\bibitem[{Troldborg et~al.(2011)Troldborg, Larsen, Madsen, Hansen, S{\o}rensen,
  and Mikkelsen}]{TroldborgEtAl2011}
Troldborg, N., Larsen, G.~C., Madsen, H.~A., Hansen, K.~S., S{\o}rensen, J.~N.,
  Mikkelsen, R., 2011. Numerical simulations of wake interaction between two
  wind turbines at various inflow conditions. Wind Energy 14~(7), 859--876.
\newline\urlprefix\url{http://dx.doi.org/10.1002/we.433}

\bibitem[{Troldborg et~al.(2010)Troldborg, S{\o}rensen, and
  Mikkelsen}]{TroldborgEtAl2010}
Troldborg, N., S{\o}rensen, J.~N., Mikkelsen, R., 2010. Numerical simulations
  of wake characteristics of a wind turbine in uniform inflow. Wind Energy
  13~(1), 86--99.
\newline\urlprefix\url{http://dx.doi.org/10.1002/we.345}

\bibitem[{Troldborg et~al.(2009)Troldborg, S{\o}rensen, and
  Mikkelsen}]{TroldborgEtAl2009}
Troldborg, N., S{\o}rensen, J.~N., Mikkelsen, R.~F., 2009. Actuator line
  modeling of wind turbine wakes. Ph.D. thesis, DTU.

\bibitem[{van Kuik et~al.(2016)van Kuik, Peinke, Nijssen, Lekou, Mann,
  S{\o}rensen, Ferreira, Vanwingerden, Schlipf, Gebraad, Polinder, Abrahamsen,
  Vanbussel, S{\o}rensen, Tavner, Bottasso, Muskulus, Matha, Lindeboom,
  Degraer, Kramer, Lehnhoff, Sonnenschein, S{\o}rensen, Künneke, Morthorst,
  and Skytte}]{vanKuikEtAl2016}
van Kuik, G. A.~M., Peinke, J., Nijssen, R., Lekou, D., Mann, J., S{\o}rensen,
  J.~N., Ferreira, C., Vanwingerden, J.~W., Schlipf, D., Gebraad, P., Polinder,
  H., Abrahamsen, A., Vanbussel, G. J.~W., S{\o}rensen, J.~D., Tavner, P.,
  Bottasso, C.~L., Muskulus, M., Matha, D., Lindeboom, H.~J., Degraer, S.,
  Kramer, O., Lehnhoff, S., Sonnenschein, M., S{\o}rensen, P.~E., Künneke,
  R.~W., Morthorst, P.~E., Skytte, K., February 2016. Long-term research
  challenges in wind energy – a research agenda by the european academy of
  wind energy. Wind Energy Science 1~(1), 1--39.
\newline\urlprefix\url{https://doaj.org/article/e57645507ee84c799ad4434a33db8ce2}

\bibitem[{Vermeer et~al.(2003)Vermeer, S{\o}rensen, and
  Crespo}]{VermeerEtAl2003}
Vermeer, L., S{\o}rensen, J., Crespo, A., 2003. Wind turbine wake aerodynamics.
  Progress in Aerospace Sciences 39~(6), 467 -- 510.
\newline\urlprefix\url{http://www.sciencedirect.com/science/article/pii/S0376042103000782}

\bibitem[{Widnall(1972)}]{Widnall1972}
Widnall, S.~E., 8 1972. The stability of a helical vortex filament. Journal of
  Fluid Mechanics 54, 641--663.

\bibitem[{Zang et~al.(1993)Zang, Street, and Koseff}]{ZangEtAl1993}
Zang, Y., Street, R.~L., Koseff, J.~R., 1993. A dynamic mixed subgrid‐scale
  model and its application to turbulent recirculating flows. Physics of Fluids
  A: Fluid Dynamics 5~(12), 3186--3196.
\newline\urlprefix\url{https://doi.org/10.1063/1.858675}

\end{thebibliography}
\end{document}